\documentclass[conference,10pt,letterpaper,nofonttune]{IEEEtran}
\IEEEoverridecommandlockouts
\usepackage{cite}
\usepackage{etoolbox}  
\usepackage{everypage} 
\usepackage{csquotes}
\usepackage{amsmath,amssymb,amsfonts}
\usepackage{algorithmic,algorithm}
\usepackage{xcolor}
\usepackage{fix-cm}
\usepackage{comment}
\usepackage{xspace}
\usepackage{balance}
\usepackage{fancyhdr}
\usepackage{bm} 
\usepackage{siunitx}

\newcommand{\mycomment}[1]{}

\newcommand{\ariadne}{ARIADNE\xspace}
\newcommand{\ack}{ACK\xspace}
\newcommand{\nack}{NACK\xspace}
\newcommand{\sionnasys}{SIONNA-SYS\xspace}
\newcommand{\sionna}{SIONNA\xspace}
\newcommand{\nvidia}{NVIDIA\xspace}
\newcommand{\nokia}{Nokia\xspace}
\usepackage{url}

\usepackage{soul}
\usepackage{multirow}
\usepackage{booktabs}
\usepackage{tabularx}
\usepackage{tikz}
\usetikzlibrary{calc}
\usepackage{tikzpagenodes}
\AddEverypageHook{%
\ifnumequal{\thepage}{1}{%
 \tikz[remember picture,overlay]{%
     \node[draw,
     text width=0.95\textwidth,
     font=\footnotesize
     ]
     at ($(current page header area) - (0,5pt)$)
     {%
   This paper has been accepted for publication at European Wireless 2026. This is the authors' accepted version of the article. The final version published by IEEE is: M. Tsampazi, N. N. Santhi, N. Perrotta, F. Dressler, T. Melodia, ``ARIADNE: AI-RAN Informed Link Adaptation in Digital Twin Network Environments,'' in Proc. of European Wireless, Rimini, Italy, June 2026.
     };
 }%
}{}
}
\usetikzlibrary{patterns, patterns.meta}

\usepackage{pgfplots}
\usepgfplotslibrary{groupplots}
\pgfplotsset{compat=1.18}
\usepackage{adjustbox}
\pgfplotsset{compat=newest}
\pgfplotsset{plot coordinates/math parser=false}
\pgfplotsset{
    every axis/.style={
        tick label style={font=\large},
        label style={font=\large},
        legend style={font=\large},
        title style={font=\large},
    }
}
\newlength\fheight
\newlength\fwidth
\usetikzlibrary{plotmarks,patterns,decorations.pathreplacing,backgrounds,calc,arrows,arrows.meta,spy,matrix,scopes,positioning, fit}
\usepgfplotslibrary{patchplots,groupplots}
\usepackage{tikzscale}
\usepackage[draft]{hyperref}

\newif\ifexttikz
\exttikzfalse

\ifexttikz
	\usetikzlibrary{external}
	\usepackage{fontspec}
\fi

\usepackage[acronyms,nonumberlist,nopostdot,nomain,nogroupskip,acronymlists={hidden}]{glossaries}
\newglossary[algh]{hidden}{acrh}{acnh}{Hidden Acronyms}
\newacronym{3gpp}{3GPP}{3rd Generation Partnership Project}
\newacronym{4g}{4G}{4th generation}
\newacronym{5g}{5G}{5th generation}
\newacronym{6g}{6G}{6th generation}
\newacronym{5gc}{5GC}{5G Core}
\newacronym{adc}{ADC}{Analog to Digital Converter}
\newacronym{aerpaw}{AERPAW}{Aerial Experimentation and Research Platform for Advanced Wireless}
\newacronym{ai}{AI}{Artificial Intelligence}
\newacronym{aimd}{AIMD}{Additive Increase Multiplicative Decrease}
\newacronym{am}{AM}{Acknowledged Mode}
\newacronym{amc}{AMC}{Adaptive Modulation and Coding}
\newacronym{amf}{AMF}{Access and Mobility Management Function}
\newacronym{aops}{AOPS}{Adaptive Order Prediction Scheduling}
\newacronym{api}{API}{Application Programming Interface}
\newacronym{xapp}{xApp}{Intelligent Application}
\newacronym{apn}{APN}{Access Point Name}
\newacronym{aqm}{AQM}{Active Queue Management}
\newacronym{ausf}{AUSF}{Authentication Server Function}
\newacronym{avc}{AVC}{Advanced Video Coding}
\newacronym{awgn}{AGWN}{Additive White Gaussian Noise}
\newacronym{balia}{BALIA}{Balanced Link Adaptation Algorithm}
\newacronym{bbu}{BBU}{Base Band Unit}
\newacronym{bdp}{BDP}{Bandwidth-Delay Product}
\newacronym{ber}{BER}{Bit Error Rate}
\newacronym{bf}{BF}{Beamforming}
\newacronym{bler}{BLER}{Block Error Rate}
\newacronym{brr}{BRR}{Bayesian Ridge Regressor}
\newacronym{bs}{BS}{Base Station}
\newacronym{bsr}{BSR}{Buffer Status Report}
\newacronym{bss}{BSS}{Business Support System}
\newacronym{ca}{CA}{Carrier Aggregation}
\newacronym{caas}{CaaS}{Connectivity-as-a-Service}
\newacronym{cb}{CB}{Code Block}
\newacronym{cc}{CC}{Congestion Control}
\newacronym{ccid}{CCID}{Congestion Control ID}
\newacronym{cv2x}{C-V2X}{Cellular Vehicle-to-Everything}
\newacronym{cco}{CC}{Carrier Component}
\newacronym{cdd}{CDD}{Cyclic Delay Diversity}
\newacronym{cdf}{CDF}{Cumulative Distribution Function}
\newacronym{cdn}{CDN}{Content Distribution Network}
\newacronym{cn}{CN}{Core Network}
\newacronym{codel}{CoDel}{Controlled Delay Management}
\newacronym{comac}{COMAC}{Converged Multi-Access and Core}
\newacronym{cord}{CORD}{Central Office Re-architected as a Datacenter}
\newacronym{cornet}{CORNET}{COgnitive Radio NETwork}
\newacronym{cosmos}{COSMOS}{Cloud Enhanced Open Software Defined Mobile Wireless Testbed for City-Scale Deployment}
\newacronym{cots}{COTS}{Commercial Off-the-Shelf}
\newacronym{cp}{CP}{Control Plane}
\newacronym{cpu}{CPU}{Central Processing Unit}
\newacronym{cqi}{CQI}{Channel Quality Information}
\newacronym{cr}{CR}{Cognitive Radio}
\newacronym{cran}{CRAN}{Cloud \gls{ran}}
\newacronym{crs}{CRS}{Cell Reference Signal}
\newacronym{csi}{CSI}{Channel State Information}
\newacronym{csirs}{CSI-RS}{Channel State Information - Reference Signal}
\newacronym{cu}{CU}{Central Unit}
\newacronym{d2tcp}{D$^2$TCP}{Deadline-aware Data center TCP}
\newacronym{d3}{D$^3$}{Deadline-Driven Delivery}
\newacronym{dac}{DAC}{Digital to Analog Converter}
\newacronym{dag}{DAG}{Directed Acyclic Graph}
\newacronym{das}{DAS}{Distributed Antenna System}
\newacronym{dash}{DASH}{Dynamic Adaptive Streaming over HTTP}
\newacronym{dc}{DC}{Dual Connectivity}
\newacronym{dccp}{DCCP}{Datagram Congestion Control Protocol}
\newacronym{dce}{DCE}{Direct Code Execution}
\newacronym{dci}{DCI}{Downlink Control Information}
\newacronym{dctcp}{DCTCP}{Data Center TCP}
\newacronym{dl}{DL}{Downlink}
\newacronym{dmr}{DMR}{Deadline Miss Ratio}
\newacronym{dmrs}{DMRS}{DeModulation Reference Signal}
\newacronym{drlcc}{DRL-CC}{Deep Reinforcement Learning Congestion Control}
\newacronym{drs}{DRS}{Discovery Reference Signal}
\newacronym{du}{DU}{Distributed Unit}
\newacronym{e2e}{E2E}{end-to-end}
\newacronym{ecaas}{ECaaS}{Edge-Cloud-as-a-Service}
\newacronym{ecn}{ECN}{Explicit Congestion Notification}
\newacronym{edf}{EDF}{Earliest Deadline First}
\newacronym{embb}{eMBB}{Enhanced Mobile Broadband}
\newacronym{empower}{EMPOWER}{EMpowering transatlantic PlatfOrms for advanced WirEless Research}
\newacronym{enb}{eNB}{evolved Node Base}
\newacronym{endc}{EN-DC}{E-UTRAN-\gls{nr} \gls{dc}}
\newacronym{epc}{EPC}{Evolved Packet Core}
\newacronym{eps}{EPS}{Evolved Packet System}
\newacronym{es}{ES}{Edge Server}
\newacronym{etsi}{ETSI}{European Telecommunications Standards Institute}
\newacronym[firstplural=Estimated Times of Arrival (ETAs)]{eta}{ETA}{Estimated Time of Arrival}
\newacronym{eutran}{E-UTRAN}{Evolved Universal Terrestrial Access Network}
\newacronym{faas}{FaaS}{Function-as-a-Service}
\newacronym{fapi}{FAPI}{Functional Application Platform Interface}
\newacronym{fdd}{FDD}{Frequency Division Duplexing}
\newacronym{fdm}{FDM}{Frequency Division Multiplexing}
\newacronym{fdma}{FDMA}{Frequency Division Multiple Access}
\newacronym{fed4fire}{FED4FIRE+}{Federation 4 Future Internet Research and Experimentation Plus}
\newacronym{olla}{OLLA}{Outer Loop Link Adaptation}
\newacronym{mdp}{MDP}{Markov Decision Process}
\newacronym{fir}{FIR}{Finite Impulse Response}
\newacronym{cir}{CIR}{Channel Impulse Response}
\newacronym{fit}{FIT}{Future \acrlong{iot}}
\newacronym{fpga}{FPGA}{Field Programmable Gate Array}
\newacronym{fr2}{FR2}{Frequency Range 2}
\newacronym{fr1}{FR1}{Frequency Range 1}
\newacronym{fs}{FS}{Fast Switching}
\newacronym{fscc}{FSCC}{Flow Sharing Congestion Control}
\newacronym{ftp}{FTP}{File Transfer Protocol}
\newacronym{fw}{FW}{Flow Window}
\newacronym{ge}{GE}{Gaussian Elimination}
\newacronym{lmmse}{LMMSE}{Linear Minimum Mean Square Error}
\newacronym{gnb}{gNB}{Next Generation Node Base}
\newacronym{nextg}{NextG}{Next Generation}
\newacronym{gop}{GOP}{Group of Pictures}
\newacronym{gpr}{GPR}{Gaussian Process Regressor}
\newacronym{gpu}{GPU}{Graphics Processing Unit}
\newacronym{gtp}{GTP}{GPRS Tunneling Protocol}
\newacronym{gtpc}{GTP-C}{GPRS Tunnelling Protocol Control Plane}
\newacronym{sca}{SCA}{Successive Convex Approximation}
\newacronym{gtpu}{GTP-U}{GPRS Tunnelling Protocol User Plane}
\newacronym{gtpv2c}{GTPv2-C}{\gls{gtp} v2 - Control}
\newacronym{gw}{GW}{Gateway}
\newacronym{harq}{HARQ}{Hybrid Automatic Repeat reQuest}
\newacronym{hetnet}{HetNet}{Heterogeneous Network}
\newacronym{hh}{HH}{Hard Handover}
\newacronym{hol}{HOL}{Head-of-Line}
\newacronym{hqf}{HQF}{Highest-quality-first}
\newacronym{hss}{HSS}{Home Subscription Server}
\newacronym{http}{HTTP}{HyperText Transfer Protocol}
\newacronym{ia}{IA}{Initial Access}
\newacronym{iab}{IAB}{Integrated Access and Backhaul}
\newacronym{ic}{IC}{Incident Command}
\newacronym{ietf}{IETF}{Internet Engineering Task Force}
\newacronym{imsi}{IMSI}{International Mobile Subscriber Identity}
\newacronym{imt}{IMT}{International Mobile Telecommunication}
\newacronym{iot}{IoT}{Internet of Things}
\newacronym{ip}{IP}{Internet Protocol}
\newacronym{itu}{ITU}{International Telecommunication Union}
\newacronym{kpi}{KPI}{Key Performance Indicator}
\newacronym{kpm}{KPM}{Key Performance Measurement}
\newacronym{kvm}{KVM}{Kernel-based Virtual Machine}
\newacronym{los}{LOS}{Line-of-Sight}
\newacronym{lsm}{LSM}{Link-to-System Mapping}
\newacronym{lstm}{LSTM}{Long Short Term Memory}
\newacronym{lte}{LTE}{Long Term Evolution}
\newacronym{lxc}{LXC}{Linux Container}
\newacronym{m2m}{M2M}{Machine to Machine}
\newacronym{mac}{MAC}{Medium Access Control}
\newacronym{manet}{MANET}{Mobile Ad Hoc Network}
\newacronym{mano}{MANO}{Management and Orchestration}
\newacronym{mc}{MC}{Multi-Connectivity}
\newacronym{mcc}{MCC}{Mobile Cloud Computing}
\newacronym{mchem}{MCHEM}{Massive Channel Emulator}
\newacronym{mcs}{MCS}{Modulation and Coding Scheme}
\newacronym{mec}{MEC}{Multi-access Edge Computing}
\newacronym{mec2}{MEC}{Mobile Edge Cloud}
\newacronym{mfc}{MFC}{Mobile Fog Computing}
\newacronym{mgen}{MGEN}{Multi-Generator}
\newacronym{mi}{MI}{Mutual Information}
\newacronym{mib}{MIB}{Master Information Block}
\newacronym{miesm}{MIESM}{Mutual Information Based Effective SINR}
\newacronym{mimo}{MIMO}{Multiple Input, Multiple Output}
\newacronym{ml}{ML}{Machine Learning}
\newacronym{mlr}{MLR}{Maximum-local-rate}
\newacronym[plural=\gls{mme}s,firstplural=Mobility Management Entities (MMEs)]{mme}{MME}{Mobility Management Entity}
\newacronym{mmtc}{mMTC}{Massive Machine-Type Communications}
\newacronym{mmwave}{mmWave}{millimeter wave}
\newacronym{mpdccp}{MP-DCCP}{Multipath Datagram Congestion Control Protocol}
\newacronym{mptcp}{MPTCP}{Multipath TCP}
\newacronym{mr}{MR}{Maximum Rate}
\newacronym{mrdc}{MR-DC}{Multi \gls{rat} \gls{dc}}
\newacronym{mse}{MSE}{Mean Square Error}
\newacronym{mss}{MSS}{Maximum Segment Size}
\newacronym{mt}{MT}{Mobile Termination}
\newacronym{mtd}{MTD}{Machine-Type Device}
\newacronym{mtu}{MTU}{Maximum Transmission Unit}
\newacronym{mumimo}{MU-MIMO}{Multi-user \gls{mimo}}
\newacronym{mvno}{MVNO}{Mobile Virtual Network Operator}
\newacronym{nalu}{NALU}{Network Abstraction Layer Unit}
\newacronym{nas}{NAS}{Non-Access Stratum}
\newacronym{nbiot}{NB-IoT}{Narrow Band IoT}
\newacronym{nfv}{NFV}{Network Function Virtualization}
\newacronym{nfvi}{NFVI}{Network Function Virtualization Infrastructure}
\newacronym{nic}{NIC}{Network Interface Card}
\newacronym{nlos}{NLOS}{Non-Line-of-Sight}
\newacronym{now}{NOW}{Non Overlapping Window}
\newacronym{nsm}{NSM}{Network Service Mesh}
\newacronym[type=hidden]{nr}{NR}{New Radio}
\newacronym{nrf}{NRF}{Network Repository Function}
\newacronym{nsa}{NSA}{Non Stand Alone}
\newacronym{nse}{NSE}{Network Slicing Engine}
\newacronym{nssf}{NSSF}{Network Slice Selection Function}
\newacronym{o2i}{O2I}{Outdoor to Indoor}
\newacronym{oai}{OAI}{OpenAirInterface}
\newacronym{oaicn}{OAI-CN}{\gls{oai} \acrlong{cn}}
\newacronym{oairan}{OAI-RAN}{\acrlong{oai} \acrlong{ran}}
\newacronym{oam}{OAM}{Operations, Administration and Maintenance}
\newacronym{ofdm}{OFDM}{Orthogonal Frequency Division Multiplexing}
\newacronym{olia}{OLIA}{Opportunistic Linked Increase Algorithm}
\newacronym{omec}{OMEC}{Open Mobile Evolved Core}
\newacronym{onap}{ONAP}{Open Network Automation Platform}
\newacronym{onf}{ONF}{Open Networking Foundation}
\newacronym{onos}{ONOS}{Open Networking Operating System}
\newacronym{oom}{OOM}{\gls{onap} Operations Manager}
\newacronym{opnfv}{OPNFV}{Open Platform for \gls{nfv}}
\newacronym{orbit}{ORBIT}{Open-Access Research Testbed for Next-Generation Wireless Networks}
\newacronym{os}{OS}{Operating System}
\newacronym{oss}{OSS}{Operations Support System}
\newacronym{pa}{PA}{Position-aware}
\newacronym{pase}{PASE}{Prioritization, Arbitration, and Self-adjusting Endpoints}
\newacronym{pawr}{PAWR}{Platforms for Advanced Wireless Research}
\newacronym{pbch}{PBCH}{Physical Broadcast Channel}
\newacronym{pcef}{PCEF}{Policy and Charging Enforcement Function}
\newacronym{pcfich}{PCFICH}{Physical Control Format Indicator Channel}
\newacronym{pcrf}{PCRF}{Policy and Charging Rules Function}
\newacronym{pdcch}{PDCCH}{Physical Downlink Control Channel}
\newacronym{pdcp}{PDCP}{Packet Data Convergence Protocol}
\newacronym{pdsch}{PDSCH}{Physical Downlink Shared Channel}
\newacronym{pdu}{PDU}{Packet Data Unit}
\newacronym{pf}{PF}{Proportionally Fair}
\newacronym{pgw}{PGW}{Packet Gateway}
\newacronym{phich}{PHICH}{Physical Hybrid ARQ Indicator Channel}
\newacronym{phy}{PHY}{Physical}
\newacronym{pmch}{PMCH}{Physical Multicast Channel}
\newacronym{pmi}{PMI}{Precoding Matrix Indicators}
\newacronym{powder}{POWDER}{Platform for Open Wireless Data-driven Experimental Research}
\newacronym{ppo}{PPO}{Proximal Policy Optimization}
\newacronym{ppp}{PPP}{Poisson Point Process}
\newacronym{prach}{PRACH}{Physical Random Access Channel}
\newacronym{prb}{PRB}{Physical Resource Block}
\newacronym{rbg}{RBG}{Resource Block Group}
\newacronym{psnr}{PSNR}{Peak Signal to Noise Ratio}
\newacronym{pss}{PSS}{Primary Synchronization Signal}
\newacronym{pucch}{PUCCH}{Physical Uplink Control Channel}
\newacronym{pusch}{PUSCH}{Physical Uplink Shared Channel}
\newacronym{qam}{QAM}{Quadrature Amplitude Modulation}
\newacronym{qci}{QCI}{\gls{qos} Class Identifier}
\newacronym{qoe}{QoE}{Quality of Experience}
\newacronym{qos}{QoS}{Quality of Service}
\newacronym{quic}{QUIC}{Quick UDP Internet Connections}
\newacronym{rach}{RACH}{Random Access Channel}
\newacronym{ran}{RAN}{Radio Access Network}
\newacronym[firstplural=end to endcess Technologies (RATs)]{rat}{RAT}{end to endcess Technology}
\newacronym{rcn}{RCN}{Research Coordination Network}
\newacronym{rec}{REC}{Radio Edge Cloud}
\newacronym{ra}{RA}{Resource Allocation}
\newacronym{red}{RED}{Random Early Detection}
\newacronym{renew}{RENEW}{Reconfigurable Eco-system for Next-generation End-to-end Wireless}
\newacronym{rfr}{RF}{Radio Frequency}
\newacronym{rfc}{RFC}{Request for Comments}
\newacronym{ric}{RIC}{\gls{ran} Intelligent Controller}
\newacronym{rlc}{RLC}{Radio Link Control}
\newacronym{rlf}{RLF}{Radio Link Failure}
\newacronym{rlnc}{RLNC}{Random Linear Network Coding}
\newacronym{rmr}{RMR}{RIC Message Router}
\newacronym{rmse}{RMSE}{Root Mean Squared Error}
\newacronym{rnis}{RNIS}{Radio Network Information Service}
\newacronym{rr}{RR}{Round Robin}
\newacronym{rrc}{RRC}{Radio Resource Control}
\newacronym{rrm}{RRM}{Radio Resource Management}
\newacronym{rru}{RRU}{Remote Radio Unit}
\newacronym{rs}{RS}{Remote Server}
\newacronym{rsrp}{RSRP}{Reference Signal Received Power}
\newacronym{rsrq}{RSRQ}{Reference Signal Received Quality}
\newacronym{rss}{RSS}{Received Signal Strength}
\newacronym{rssi}{RSSI}{Received Signal Strength Indicator}
\newacronym{rtt}{RTT}{Round Trip Time}
\newacronym{ru}{RU}{Radio Unit}
\newacronym{rw}{RW}{Receive Window}
\newacronym{rx}{RX}{Receiver}
\newacronym{s1ap}{S1AP}{S1 Application Protocol}
\newacronym{sa}{SA}{standalone}
\newacronym{sack}{SACK}{Selective Acknowledgment}
\newacronym{sap}{SAP}{Service Access Point}
\newacronym{sc2}{SC2}{Spectrum Collaboration Challenge}
\newacronym{scef}{SCEF}{Service Capability Exposure Function}
\newacronym{sch}{SCH}{Secondary Cell Handover}
\newacronym{scoot}{SCOOT}{Split Cycle Offset Optimization Technique}
\newacronym{sctp}{SCTP}{Stream Control Transmission Protocol}
\newacronym{sdap}{SDAP}{Service Data Adaptation Protocol}
\newacronym{sdk}{SDK}{Software Development Kit}
\newacronym{sdm}{SDM}{Space Division Multiplexing}
\newacronym{sdma}{SDMA}{Spatial Division Multiple Access}
\newacronym{sdn}{SDN}{Software-defined Networking}
\newacronym{sdr}{SDR}{Software-defined Radio}
\newacronym{seba}{SEBA}{SDN-Enabled Broadband Access}
\newacronym{sgsn}{SGSN}{Serving GPRS Support Node}
\newacronym{sgw}{SGW}{Service Gateway}
\newacronym{si}{SI}{Study Item}
\newacronym{sib}{SIB}{Secondary Information Block}
\newacronym{sinr}{SINR}{Signal to Interference plus Noise Ratio}
\newacronym{sip}{SIP}{Session Initiation Protocol}
\newacronym{siso}{SISO}{Single Input, Single Output}
\newacronym{sla}{SLA}{Service Level Agreement}
\newacronym{sm}{SM}{Service Model}
\newacronym{smf}{SMF}{Session Management Function}
\newacronym{smo}{SMO}{Service Management and Orchestration}
\newacronym{sms}{SMS}{Short Message Service}
\newacronym{smsgmsc}{SMS-GMSC}{\gls{sms}-Gateway}
\newacronym{snr}{SNR}{Signal-to-Noise-Ratio}
\newacronym{son}{SON}{Self-Organizing Network}
\newacronym{sptcp}{SPTCP}{Single Path TCP}
\newacronym{srb}{SRB}{Service Radio Bearer}
\newacronym{srn}{SRN}{Standard Radio Node}
\newacronym{srs}{SRS}{Sounding Reference Signal}
\newacronym{ss}{SS}{Synchronization Signal}
\newacronym{sss}{SSS}{Secondary Synchronization Signal}
\newacronym{st}{ST}{Spanning Tree}
\newacronym{svc}{SVC}{Scalable Video Coding}
\newacronym{tb}{TB}{Transport Block}
\newacronym{tcp}{TCP}{Transmission Control Protocol}
\newacronym{tdd}{TDD}{Time Division Duplexing}
\newacronym{tdm}{TDM}{Time Division Multiplexing}
\newacronym{tdma}{TDMA}{Time Division Multiple Access}
\newacronym{tfl}{TfL}{Transport for London}
\newacronym{tfrc}{TFRC}{TCP-Friendly Rate Control}
\newacronym{tft}{TFT}{Traffic Flow Template}
\newacronym{tgen}{TGEN}{Traffic Generator}
\newacronym{fqi}{FQI}{Fitted Q-Iteration}
\newacronym{tip}{TIP}{Telecom Infra Project}
\newacronym{mab}{MAB}{Multi-Armed Bandit} 
\newacronym{tm}{TM}{Transparent Mode}
\newacronym{to}{TO}{Telco Operator}
\newacronym{tr}{TR}{Technical Report}
\newacronym{trp}{TRP}{Transmitter Receiver Pair}
\newacronym{ts}{TS}{Technical Specification}
\newacronym{tti}{TTI}{Transmission Time Interval}
\newacronym{ttt}{TTT}{Time-to-Trigger}
\newacronym{tx}{TX}{Transmitter}
\newacronym{uas}{UAS}{Unmanned Aerial System}
\newacronym{uav}{UAV}{Unmanned Aerial Vehicle}
\newacronym{udm}{UDM}{Unified Data Management}
\newacronym{udp}{UDP}{User Datagram Protocol}
\newacronym{udr}{UDR}{Unified Data Repository}
\newacronym{ue}{UE}{User Equipment}
\newacronym{uhd}{UHD}{\gls{usrp} Hardware Driver}
\newacronym{ul}{UL}{Uplink}
\newacronym{ap}{AP}{Access Point}
\newacronym{um}{UM}{Unacknowledged Mode}
\newacronym{uml}{UML}{Unified Modeling Language}
\newacronym{upa}{UPA}{Uniform Planar Array}
\newacronym{upf}{UPF}{User Plane Function}
\newacronym{urllc}{URLLC}{Ultra Reliable and Low Latency Communications}
\newacronym{usa}{U.S.}{United States}
\newacronym{usim}{USIM}{Universal Subscriber Identity Module}
\newacronym{usrp}{USRP}{Universal Software Radio Peripheral}
\newacronym{utc}{UTC}{Urban Traffic Control}
\newacronym{vim}{VIM}{Virtualization Infrastructure Manager}
\newacronym{vm}{VM}{Virtual Machine}
\newacronym{vnf}{VNF}{Virtual Network Function}
\newacronym{volte}{VoLTE}{Voice over \gls{lte}}
\newacronym{voltha}{VOLTHA}{Virtual OLT HArdware Abstraction}
\newacronym{vr}{VR}{Virtual Reality}
\newacronym{vran}{vRAN}{Virtualized \gls{ran}}
\newacronym{vss}{VSS}{Video Streaming Server}
\newacronym{wbf}{WBF}{Wired Bias Function}
\newacronym{wf}{WF}{Waterfilling}
\newacronym{wlan}{WLAN}{Wireless Local Area Network}
\newacronym{osm}{OSM}{Open Source \gls{nfv} Management and Orchestration}
\newacronym{pnf}{PNF}{Physical Network Function}
\newacronym{drl}{DRL}{Deep Reinforcement Learning}
\newacronym{rl}{RL}{Reinforcement Learning}
\newacronym{mtc}{MTC}{Machine-type Communications}
\newacronym{osc}{OSC}{O-RAN Software Community}
\newacronym{rc}{RC}{RAN Control}
\newacronym{dqn}{DQN}{Deep Q-Network}
\newacronym{v2x}{V2X}{Vehicle-to-everything}
\newacronym{gbsm}{GBSM}{Geometry-Based Stochastic Model}
\newacronym{gbs}{GBSM}{Geometry-Based Stochastic}
\newacronym{quadriga}{QuaDRiGa}{QUAsi Deterministic RadIo channel GenerAtor}
\newacronym{relu}{ReLU}{Rectified Linear Unit} 
\newacronym{mpc}{MPC}{Multipath Component}
\newacronym{xpr}{XPR}{Cross-polarization Ratio}
\newacronym{lsp}{LSP}{Large Scale Parameter}
\newacronym{ssp}{SSP}{Small Scale Parameter}
\newacronym{fbs}{FBS}{First Bounce Scatterer}
\newacronym{lbs}{LBS}{Last Bounce Scatterer}
\newacronym{d2d}{D2D}{Device-to-Device}
\newacronym{rsu}{RSU}{Road Side Unit}
\newacronym{toa}{ToA}{Time-of-Arrival}
\newacronym{ris}{RIS}{Reconfigurable Intelligent Surface}
\newacronym{aoa}{AoA}{Angle of Arrival}
\newacronym{aod}{AoD}{Angle of Departure}
\newacronym{pl}{PL}{Path-Loss}
\newacronym{noma}{NOMA}{Non-Orthogonal Multiple Access}
\newacronym{sic}{SIC}{Successive Interference Cancellation}
\newacronym{gps}{GPS}{Global Positioning System}
\newacronym{ids}{IDS}{Independent Diffusive Scatterer-based}
\newacronym{inw}{INW}{Impedance Network-based}
\newacronym{minlp}{MINLP}{Mixed Integer Non-Linear Programming}
\newacronym{star}{STAR}{Simultaneous Transmitting And Reflecting}
\newacronym{leo}{LEO}{Low Earth Orbit}
\newacronym{svd}{SVD}{Singular Value Decomposition}
\newacronym{kkt}{KKT}{Karush-Kuhn-Tucker}
\newacronym{inr}{INR}{interference-to-noise ratio}
\newacronym{ota}{OTA}{over-the-air}
\newacronym{salad}{SALAD}{Self-Adaptive Link Adaptation}
\newacronym{ariadne}{ARIADNE}{AI-RAN Informed Link Adaptation in Digital Twin Network Environments}
\newacronym{se}{SE}{Spectral Efficiency}
\newacronym{dt}{DT}{Decision Tree}
\newacronym{kf}{KF}{Kalman Filter}
\newacronym{rf}{RF}{Random Forest}
\newacronym{oco}{OCO}{Online Convex Optimization}
\newacronym{dcqi}{dCQI}{delayed CQI}
\newacronym{mlp}{MLP}{Multilayer Perceptron}
\newacronym{rzf}{RZF}{Regularized Zero-Forcing}

\glsdisablehyper

\begin{document}


\title{ARIADNE: AI-RAN Informed Link Adaptation in Digital Twin Network Environments \vspace{-.3cm}}

\author{\IEEEauthorblockN{Maria Tsampazi\IEEEauthorrefmark{1}, Neagin Neasamoni Santhi\IEEEauthorrefmark{1}, Nicole Perrotta\IEEEauthorrefmark{1}, Falko Dressler\IEEEauthorrefmark{4}, Tommaso Melodia\IEEEauthorrefmark{1}}
\IEEEauthorblockA{\IEEEauthorrefmark{1}Institute for Intelligent Networked Systems, Northeastern University, Boston, MA, U.S.A.\\E-mail: \{tsampazi.m, neasamonisanthi.n, perrotta.nic, t.melodia\}@northeastern.edu\\\IEEEauthorrefmark{4}School of Electrical Engineering and Computer Science, TU Berlin, Germany\\E-mail: \{dressler\}@ccs-labs.org}
\thanks{This article is based upon work supported by the U.S.\ National Science Foundation under Grant CNS-2112471.}
}

\makeatletter
\patchcmd{\@maketitle}
  {\addvspace{0.5\baselineskip}\egroup} 
  {\addvspace{-2\baselineskip}\egroup} 
  {}
  {}
\makeatother

\maketitle

\glsunset{usrp}

\begin{abstract}
\gls{ai}-powered \gls{ran} networks have attracted 
significant attention from both industry and academia. Meanwhile, 
Digital Twins offer a safe playground for experimenting with 
\gls{ai}/\gls{ml}-based solutions for advanced \gls{ai}-\gls{ran} 
research.
By enabling the testing of online algorithms 
before deployment on the \gls{ran}, they reduce costs and safety risks 
associated with physical field testing. In this article, we propose \ariadne, an online \gls{rl}-based module that seamlessly 
integrates with \sionna and is tasked with performing link adaptation. 
We 
explore different design choices and demonstrate how \ariadne can 
surpass industry-standard and state-of-the-art methods by achieving up 
to $11\%$ and $20\%$ improvements in Spectral Efficiency, respectively.
Finally, we show that \gls{rl} learns a \gls{mcs} selection strategy that diverges from \gls{olla}, exhibiting either more conservative or more aggressive behavior depending on the configuration, a trend further corroborated by training offline on \gls{5g} \gls{ota} measurements.
\end{abstract}

\glsresetall
\glsunset{nextg}
\glsunset{ota}
\glsunset{mcs}
\glsunset{ris}
\glsunset{ran}
\glsunset{enb}
\glsunset{embb}
\glsunset{urllc}
\glsunset{ric}
\glsunset{ai}
\glsunset{ml}
\glsunset{6g}
\glsunset{5g}
\glsunset{rl}


\vspace{-0.35cm}
\section{Introduction}\label{intro}
\vspace{-0.1cm}
Recent years have seen collaboration among leading industry entities such as \nokia and \nvidia to advance \gls{ai}-\gls{ran} research~\cite{nvidia_nokia2025}. \gls{ai}-native \gls{6g} is expected to emerge from simulation, with Digital Twins playing a key role in the train--simulate--deploy--optimize lifecycle~\cite{nokia_ran_dt2026}. Indeed, Digital Twins are envisioned to enable faster innovation by allowing the evaluation of \enquote{what-if} scenarios for future technologies that are not yet available in hardware~\cite{polese2024colosseum}. By reducing the \enquote{concept-to-live} cycle, dense urban environments and complex \gls{6g} use cases can be simulated with greater accuracy, supporting the design and delivery of \gls{nextg} systems while accelerating the adoption and deployment of advanced \gls{ai} solutions~\cite{nokia_ran_dt2026}. 

In this context, \nvidia's \sionna Digital Twin environment~\cite{sionna2022} provides an all-in-one platform for wireless research, enabling the execution of system-level simulations over ray-traced channels. In particular, \sionnasys allows the on-the-fly integration of plugin \gls{ai}/\gls{ml} solutions, facilitating the testing of key features and capabilities essential to \gls{ai}-native \gls{6g} research. Simulated, controlled environments often provide a practical first step for developing \gls{ai}/\gls{ml} solutions before conducting \gls{ota} experiments, allowing for algorithm refinement and parameter optimization.

At the same time, a trending topic that has gained widespread attention in the research community is the optimization of the \gls{gnb}'s \gls{mac}-layer scheduler. This involves the optimal allocation of \glspl{prb} and selection of the \gls{mcs} index, as well as the tuning of power control mechanisms (e.g., 3GPP-compliant open and closed-loop power control) that adjust transmit power based on predefined \gls{snr} targets to meet \gls{qos} requirements for different slices and verticals. Importantly, by jointly optimizing power control and \gls{mcs} selection, Spectral Efficiency can be significantly improved, demonstrating the benefits of a coordinated approach for enhanced network management.

\vspace{-0.3cm}
\subsection{Related Work}\label{relatedwork}
\vspace{-0.12cm}
Numerous works have focused on \gls{rl} for \gls{mcs} selection~\cite{khedhri2025adaptive}, ranging from offline approaches~\cite{peri2025offline} to contextual \gls{mab} solutions~\cite{pulliyakode2017reinforcement,saxena2019contextual}. The authors in~\cite{zubow2021grgym} introduce GrGym, an \gls{rl}-based framework for \gls{mcs} selection in WiFi, which leverages a custom gym environment. In~\cite{leite2012flexible}, the authors formulate an \gls{rl} framework where a low discount factor enforces myopic \gls{mcs} selection based on the current channel state. The industry standard for link adaptation is \gls{olla}~\cite{pedersen2007frequency}, and modifications involving \gls{rl} have been introduced for intelligent on-the-fly adaptation~\cite{kela2022reinforcement}. More recently, the authors in~\cite{wiesmayr2025salad} propose a gradient descent approach with a learning rate that self-adapts online through knowledge distillation. However, none of the works mentioned above focuses on the integration of an \gls{ai} module on the fly within high-fidelity Digital Twins.

\vspace{-0.3cm}
\subsection{Contributions}\label{contributions}
\vspace{-0.12cm}
Motivated by the suitability of system-level simulations for digital twin environments, we propose \ariadne, a framework that leverages \nvidia's platform \sionna to integrate an \gls{rl}-module for \gls{ai}-\gls{ran} networks, enabling learning-based link adaptation through adaptive \gls{mcs} selection on channels simulated via \sionna ray tracing. Our module seamlessly integrates with the \sionnasys platform, enabling direct comparison with state-of-the-art link adaptation algorithms (such as the industry-standard \gls{olla}~\cite{pedersen2007frequency} and \nvidia's proposed link adaptation scheme, entitled \gls{salad}~\cite{wiesmayr2025salad}). In this article, we train and evaluate \ariadne on a variety of high-fidelity ray-traced channels generated with \sionna, and we demonstrate that online \gls{rl} outperforms both \gls{olla} and \gls{salad} in \gls{mcs} selection in terms of performance, as measured by the achieved Spectral Efficiency. Finally, we evaluate the suitability of \gls{rl} on offline \gls{5g} data collected from an \gls{ota} \gls{5g} testbed, demonstrating its effectiveness on real-world measurements. 

\vspace{-0.2cm}
\section{\ariadne: Learning Link Adaptation via Reinforcement Learning}\label{ariadne-intro}
\vspace{-0.15cm}
Our \gls{rl} agent operates at the \textit{Fast Link Adaptation} level, where decisions (i.e., the selection of \gls{mcs} levels for all \glspl{ue}) are enforced on a per-slot basis. This requires precise knowledge of the channel quality. However, \gls{cqi} reports are subject to feedback delay~\cite{maggi2026sinr}. In addition, \gls{cqi} is typically reported over a wide band, while transmission may occur over a narrower band. As a result, the true \gls{sinr} is not directly observable and must be estimated online at each slot using only past feedback. Consequently, link adaptation requires jointly inferring the channel quality and selecting an appropriate \gls{mcs}~\cite{wiesmayr2025salad}. Therefore, for the current slot's channel quality, we consider two estimation modes: an \textit{Oracle} mode that establishes an upper bound using \sionna's simulated environment to calculate the \gls{sinr}, and a predictor mode for \gls{sinr} estimation that leverages previous \gls{cqi} reports.
 
\textbf{\textit{\gls{mdp} Formulation and Temporal Resolution.}} We formulate the \gls{mcs} selection problem as an \gls{mdp} defined by the tuple $(\mathcal{S}, \mathcal{A}, R, P, \beta)$. Critical to our design is a $1$~:~$1$ mapping between the \gls{rl} environment and the \gls{5g}~\gls{nr} \gls{phy}-layer, where each discrete environment step corresponds exactly to one \gls{5g}~\gls{nr} time slot.

\vspace{-0.18cm}
\subsection{State Space}\label{state-space}

At each time slot $t$, the agent observes a state vector $\mathbf{s}_t \in \mathbb{R}^{(N_\mathrm{cqi} + N_\mathrm{harq} + 5) \cdot U}$, where $U$ is the number of \glspl{ue}. The state is composed of per-\gls{ue} feature vectors:
\begin{equation}
    \mathbf{s}_t = \left[\mathbf{s}_t^{(1)}, \ldots, \mathbf{s}_t^{(U)}\right],
\end{equation}
where each per-\gls{ue} component $\mathbf{s}_t^{(u)}$ is given by
\begin{equation}
    \mathbf{s}_t^{(u)} = \left[\tilde{\gamma}_{\mathrm{post},t}^{(u)}, \; \tilde{\gamma}_{\mathrm{eff},t-k}^{(u)}, \; h_{t-k}^{(u)},\; \tilde{m}_{t-1}^{(u)},\; \hat{B}_t^{(u)},\; \Delta_{\mathrm{offset},t}^{(u)},\; \bar{a}_t^{(u)}\right],
\end{equation}
and the respective components are defined as follows:
\begin{itemize}
    \item $\tilde{\gamma}_{\mathrm{post},t}^{(u)}$: An estimate of the current slot's normalized post-equalization \gls{sinr}, computed as the mean \gls{sinr} across allocated resource elements after \gls{lmmse} equalization. In \textit{Oracle} mode, this is derived from the current channel realization; in \textit{predictor} mode, it is estimated from past measurements. 
    \item $\tilde{\gamma}_{\mathrm{eff},t-k}^{(u)}$, $k = 1, \ldots, N_{\mathrm{cqi}}$: The $N_{\mathrm{cqi}}$ most recent effective normalized \gls{sinr} values reported by the \gls{phy}-layer Abstraction.
    \item $h_{t-k}^{(u)}$, $k = 1, \ldots, N_{\mathrm{harq}}$: A window of $N_{\mathrm{harq}}$ prior \gls{harq} outcomes, where $h \in \{-1, 0, 1\}$ denotes unscheduled, \nack, and \ack, respectively.
    \item $\tilde{m}_{t-1}^{(u)}$: The normalized previously selected \gls{mcs} index.
    \item $\hat{B}_t^{(u)}$: The sliding-window \gls{bler} estimate over the last $N_{\mathrm{bler}}$ scheduled slots, calculated as $\hat{B}_t^{(u)} = \sum \mathrm{NACK} \,/\, \sum \mathrm{Scheduled}$.
    \item $\Delta_{\mathrm{offset},t}^{(u)}$: A normalized asymmetric feedback 
accumulator updated by $+0.1$ on \ack\ and $-0.9$ on \nack.\footnotemark
    \item $\bar{a}_t^{(u)}$: The running \ack\ rate over the full 
episode.\footnotemark[\value{footnote}]
\end{itemize}
\footnotetext{Both $\Delta_{\mathrm{offset},t}^{(u)}$ and $\bar{a}_t^{(u)}$ 
are derived from \gls{harq} feedback; the former weights \nack\ outcomes 
$9\times$ more heavily than \ack, making it more sensitive to link 
failures than the symmetric \ack\ rate $\bar{a}_t^{(u)}$.}
\vspace{-0.25cm}
\subsection{Action Space}\label{action-space}
The agent performs link adaptation for the \gls{5g}~\gls{nr} \gls{pdsch}.
The action $\mathbf{a}_t \in \{0, 1, \ldots, 28\}^U$ is a vector of \gls{mcs} indices, one per \gls{ue}, as defined in the \gls{5g}~\gls{nr} \gls{mcs} Table~$1$~\cite{3gpp38214}. The mapping is performed directly by the policy network as follows in~\eqref{mapping}:
\begin{equation}\label{mapping}
    \mathbf{a}_t = \pi(\mathbf{s}_t) = \left(m_t^{(1)}, \ldots, m_t^{(U)}\right).
\end{equation}

\vspace{-0.5cm}
\subsection{Reward}\label{rewarddef}
\vspace{-0.1cm}
We aim to capture the fundamental \gls{mcs} selection tradeoff. Higher \gls{mcs} indices result in higher Spectral Efficiency, but simultaneously increase the \gls{bler} and, consequently, the probability of decoding failure. Conversely, a conservative \gls{mcs} lowers the \gls{bler} but also reduces the Spectral Efficiency. To target the practical throughput--reliability tradeoff, we maximize a reward that captures the effective Spectral Efficiency, which is directly proportional to system goodput.

\textbf{\textit{Instantaneous Reward.}} At each environment step (i.e., each \gls{5g}~\gls{nr} time slot), the agent selects an \gls{mcs} index for each \gls{ue}, and the environment subsequently returns a scalar reward based on the successfully delivered Spectral Efficiency in that slot. The instantaneous reward $r_t$ at slot $t$ is defined as the total achieved Spectral Efficiency across all \glspl{ue}, as expressed in~\eqref{eq:reward}:

\begin{equation}
    r_t = \sum_{u=1}^{U} Q(m_t^{(u)}) \cdot R_c(m_t^{(u)}) \cdot \mathbb{1} \left\{ \mathcal{H}_t^{(u)} = \mathrm{ACK} \right\},
    \label{eq:reward}
\end{equation}

\noindent
where $Q(m_t^{(u)})$ and $R_c(m_t^{(u)})$ denote the modulation order and the effective code rate, respectively, associated with the \gls{mcs} index $m_t^{(u)}$ selected for \gls{ue} $u$ at slot $t$. The term $\mathbb{1}\{\cdot\}$ denotes the indicator function, which evaluates to $1$ if the \gls{harq} outcome $\mathcal{H}_t^{(u)}$ for the $u$-th \gls{ue} is a successful acknowledgment ($\mathrm{ACK}$), and $0$ otherwise.

\textbf{\textit{Expected Reward.}} In the \sionnasys simulator, the \gls{harq} feedback is generated by the \gls{phy} Abstraction module as a Bernoulli outcome with success probability $\Pr(\mathrm{ACK}_t^{(u)}=1\mid\mathbf{s}_t,a_t)=1-\mathrm{BLER}(m_t^{(u)},\gamma_t^{(u)})$, where the \gls{bler} depends on the selected \gls{mcs} index $m_t^{(u)}$ and the effective \gls{sinr} $\gamma_t^{(u)}$ computed from the post-equalization \gls{sinr} at slot $t$. 
\noindent
Given that the nominal Spectral Efficiency is defined as $\mathrm{SE}^{(u)}_{\mathrm{nom},t} \triangleq Q(m_t^{(u)})\cdot R_c(m_t^{(u)})$ for a chosen \gls{mcs}, the expected per-slot reward is given in~\eqref{eq:expected_reward_final}:
\begin{equation}
\mathbb{E}[r_t\mid\mathbf{s}_t, a_t]=\sum_{u=1}^{U}
      \underbrace{Q(m_t^{(u)}) \cdot R_c(m_t^{(u)})}_{
        \text{SE}_{\text{nom}}(m_{t})}
      \cdot \bigl(1-\text{\small{BLER}}(m_t^{(u)}, \gamma_t^{(u)})\bigr).
  \label{eq:expected_reward_final}
\end{equation}

\begin{figure}[h]
\centering
\includegraphics[width=1.05\columnwidth, height=6cm, keepaspectratio]{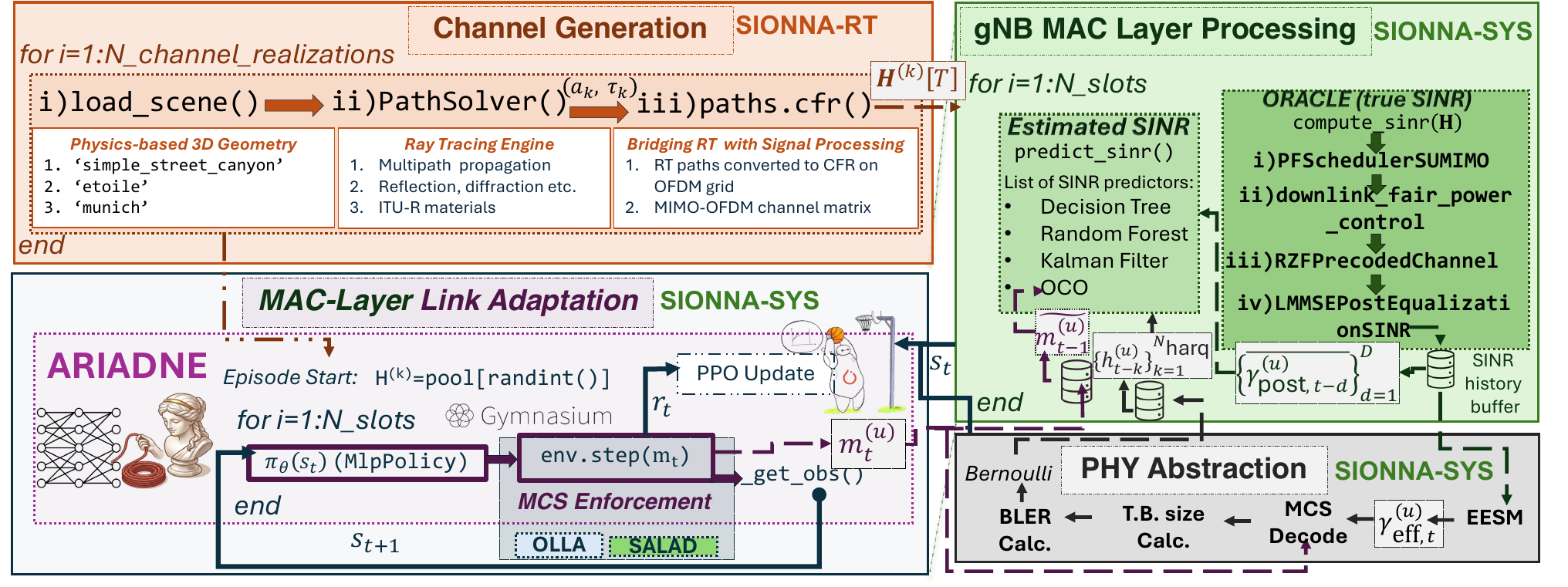}
\setlength{\abovecaptionskip}{-12pt}
\setlength{\belowcaptionskip}{-15pt}
\caption{End-to-end execution of \gls{mcs} selection in \sionna. The channel matrix is generated from a pool of ray-traced channel realizations to ensure diversity. The current 
\gls{sinr} is either directly used or estimated from previous observations. 
The post-equalization \gls{sinr} is then provided as input to the 
\gls{phy} Abstraction. The state $\mathbf{s}_t$ is observed by 
the \ariadne\ framework, which selects the corresponding \gls{mcs} index.}
\label{pipeline_total}
\vspace{-0.78cm}
\end{figure}

\vspace{-0.1cm}
\textbf{\textit{Reward Interpretation.}} The expectation in~\eqref{eq:expected_reward_final} evaluates the effective Spectral Efficiency, which is directly proportional to the system goodput. Our \gls{rl} agent is tasked with maximizing the immediate per-slot reward, which over an episode is equivalent to maximizing the time-average effective throughput. Therefore, our reward formulation naturally captures the fundamental link adaptation tradeoff, with the \gls{rl} agent learning a policy that selects the \gls{mcs} to maximize the product $\mathrm{SE}_{\mathrm{nom}}\cdot(1 - \mathrm{BLER})$, thereby learning the rate--reliability tradeoff through trial and error.

\textbf{\textit{Adaptive \gls{bler} Penalty.}} Although the reward 
in~\eqref{eq:reward} implicitly penalizes \gls{bler} through zero Spectral Efficiency on \nack, this signal alone may not be sufficient to prevent persistently \textit{aggressive} \gls{mcs} selection. Link adaptation mechanisms such as \gls{olla} target a fixed \gls{bler} 
level (typically around $10\%$~\cite{3gpp38214,wiesmayr2025salad,pedersen2007frequency}), balancing throughput and reliability. To incorporate a similar notion of 
reliability control, we additionally augment the reward with a per-\gls{ue} penalty weight $\lambda_t^{(u)}$, governed by an integral controller that increases when the cumulative \gls{bler} exceeds a target $\tau$ and remains zero otherwise, as defined in~\eqref{eq:penaltyreward}--\eqref{eq:explicit-penalty-reward}:
\vspace{-0.25cm}
\begin{equation}
r_t=\sum_{u}\Big[\mathrm{SE}_{\mathrm{nom},t}^{(u)}\mathbb{1}\{\mathcal{H}_t^{(u)}=\mathrm{ACK}\}
-\lambda_t^{(u)}\mathbb{1}\{\mathcal{H}_t^{(u)}=\mathrm{NACK}\}\Big],
\label{eq:penaltyreward}
\end{equation}
\noindent
\vspace{-0.25cm}
where
\begin{equation}
\lambda_t^{(u)}=\max\!\Big[0,\;k_E\!\!\sum_{\substack{i\le t\\ \mathcal{H}_i^{(u)}\neq\varnothing}}
\bigl(\mathbb{1}\{\mathcal{H}_i^{(u)}=\mathrm{NACK}\}-\tau\bigr)\Big].
\label{eq:explicit-penalty-reward}
\end{equation}
\noindent
When $\lambda_t^{(u)} = 0$ for all $u \in \{1,\ldots,U\}$,~\eqref{eq:penaltyreward} 
reduces to~\eqref{eq:reward}.

\vspace{-0.3cm}
\subsection{Transition Dynamics}\label{transition-dyn}
\vspace{-0.15cm}
The environment transitions $P(\mathbf{s}_{t+1} | \mathbf{s}_t, \mathbf{a}_t)$ are determined by the \sionnasys simulator as detailed in Section~\ref{systemmodel}.

\vspace{-0.35cm}
\section{System Model}\label{systemmodel}
\vspace{-0.2cm}
\begin{figure}[h]
  \centering
  \includegraphics[width=1.02\columnwidth, height=6cm]{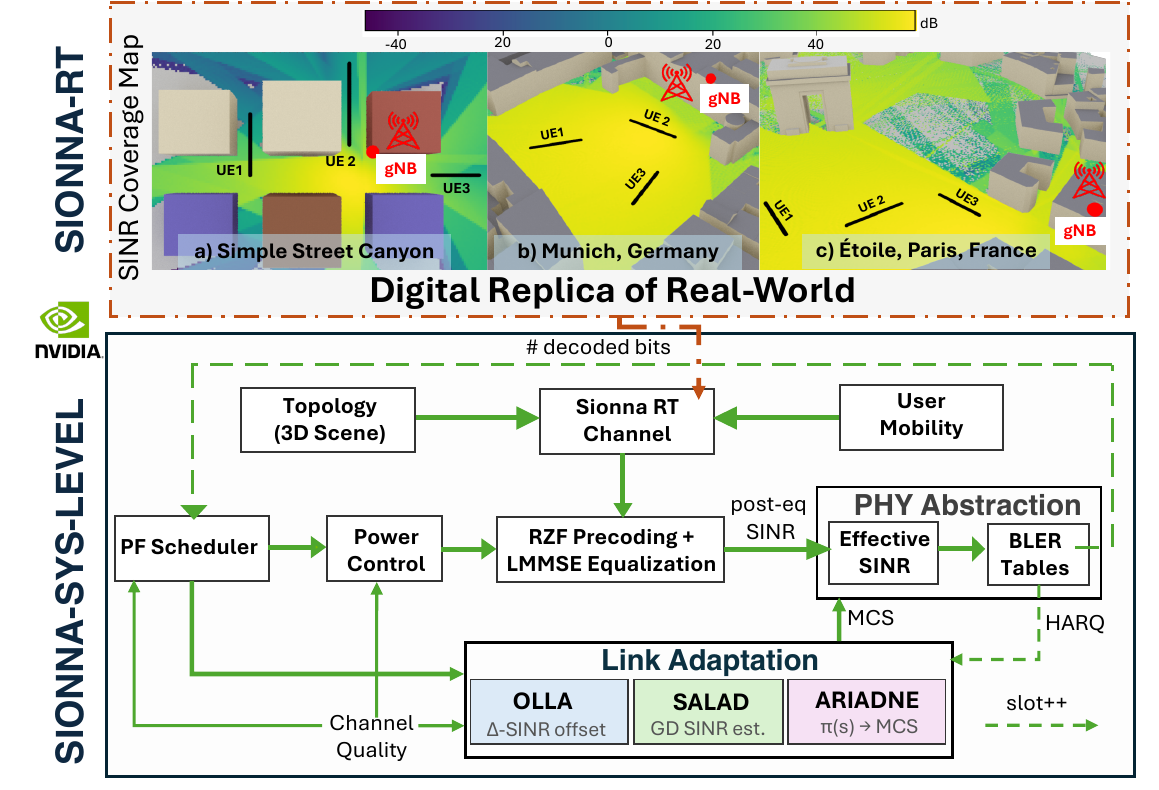}
  \setlength{\abovecaptionskip}{-10pt}
  \setlength{\belowcaptionskip}{-8pt}
  \caption{System-level simulations using \sionna over ray-traced channels for link adaptation with \gls{olla}, \ariadne, and \gls{salad}.}
  \label{fig:syslevelsim}
  \vspace{-0.7cm}
\end{figure}

We compare three link adaptation strategies, namely \gls{olla}, 
\gls{salad}, and \ariadne.
All operate within 
the same \sionnasys pipeline,
and the decision-making component resides within the \gls{gnb}'s 
\gls{mac}-layer scheduler. Independent of the specific algorithm employed, this reflects a centralized architecture in which the \gls{gnb} collects 
\gls{cqi} and \gls{harq} feedback from all \glspl{ue}, executes the link adaptation algorithm internally, and assigns \gls{mcs} indices to scheduled \glspl{ue} for the upcoming transmission. The per-slot pipeline is illustrated in 
Fig.~\ref{fig:syslevelsim}.
\vspace{-0.25cm}
\subsection{System Architecture and Link Adaptation Workflow}\label{sec:system-model-arch}
\vspace{-0.1cm}
The end-to-end pipeline is shown in Fig.~\ref{pipeline_total}.
At the beginning of each simulation, a 3D scene (e.g., Urban Street Canyon, 
Munich, or \'{E}toile in Fig.~\ref{fig:syslevelsim}) is loaded into \sionna's ray tracer to compute the 
channel frequency response $\mathbf{H}_t$ per slot. At each slot, the simulator executes proportional-fair scheduling, power control, and \gls{rzf} precoding with \gls{lmmse} equalization, yielding 
post-equalization \gls{sinr} values. Link 
adaptation then selects an \gls{mcs} index based on the \gls{sinr} and 
\gls{harq} feedback. This is the only component that differs across methods (Fig.~\ref{fig:syslevelsim}). \gls{olla} maintains a per-\gls{ue} \gls{sinr} offset updated via 
\ack/\nack feedback and selects the highest \gls{mcs} under a target 
\gls{bler}, using the latest \gls{sinr} report and \gls{harq} outcome. 
\gls{salad} primarily relies on \ack/\nack feedback, estimating the 
\gls{sinr} online via recursive updates and adapting the \gls{bler} 
target through a feedback control loop. 
\ariadne observes a window of past \gls{cqi} and \gls{harq} feedback, 
together with a predicted \gls{sinr}, and directly maps this state to 
\gls{mcs} indices via a trained policy network. 
Finally, \gls{phy} Abstraction maps the selected \gls{mcs} and 
post-equalization \gls{sinr} to a \gls{bler}, from which the \gls{harq} 
outcome is sampled, yielding the achieved Spectral Efficiency. 
All steps except link adaptation are identical across methods, ensuring that any 
performance difference is solely due to the \gls{mcs} selection strategy.

\vspace{-0.25cm}
\subsection{Implementation and Training}\label{ppo-design}
\vspace{-0.1cm}
We train the agent using \gls{ppo}~\cite{schulman2017proximal} as implemented in Stable-Baselines3~\cite{stable-baselines3} within a custom Gymnasium environment~\cite{gymnasium2023} built on top of the \sionna system-level simulator, with \ariadne handling agent--environment interaction. The policy is parameterized by a two-layer \gls{mlp} with $64$ hidden units per layer, a learning rate of $3\times10^{-4}$, a clip range of $\varepsilon=0.2$, and an entropy coefficient of $0.05$ to encourage exploration, resulting in low computational complexity suitable for \gls{gnb} deployment. 
We set the discount factor $\beta=0$, treating the problem as a \textit{contextual bandit}. This reflects the observation that \gls{mcs} selection is primarily driven by instantaneous channel state, and temporal credit assignment for future slots does not improve performance.
In time-varying fading channels, channel temporal correlation decreases rapidly, limiting the reliability of long-horizon credit assignment. A myopic policy (i.e., $\beta=0$) is therefore well suited to this setting. In our setup, we assume three mobile \glspl{ue} while the ray-traced channels correspond to \sionna's Simple Street Canyon scenario.

\vspace{-0.2cm}
\section{Experimental Evaluation}\label{expevalresults}
\vspace{-0.25cm}
\ariadne observes a configurable window of $N_{\mathrm{cqi}}$ past effective 
\gls{sinr} reports and $N_{\mathrm{harq}}$ prior \gls{harq} outcomes, as 
highlighted in Section~\ref{state-space}. We evaluate two 
configurations, as shown in Table~\ref{tab:configurations}. In Setup~A, the observation window leverages $N_{\mathrm{cqi}}\!=\!3$ 
reports and $N_{\mathrm{harq}}\!=\!10$ outcomes, providing the agent with temporal context. In Setup~B, the observation is limited to a single 
past \gls{sinr} report and a single \gls{harq} outcome. Finally, all experimental results reported below have been collected over multiple experiments and channel realizations.
\begin{table}[h]
\vspace{-0.5cm}
\centering
\setlength{\abovecaptionskip}{-2.15pt}  
\setlength{\belowcaptionskip}{-0.25pt}
\caption{Observation Size configurations for \ariadne.}
\label{tab:configurations}
\footnotesize
\begin{tabular}{@{}lcc@{}}
\toprule
\textbf{Parameter} & \textbf{Setup~A} & \textbf{Setup~B} \\
\midrule
\gls{cqi} window ($N_{\mathrm{cqi}}$)       & 3  & 1  \\
\gls{harq} window ($N_{\mathrm{harq}}$)     & 10  & 1 \\[-1.5pt]
\bottomrule
\end{tabular}
\vspace{-0.4cm}
\end{table}
\subsection{Impact of Perfect and Imperfect Channel State Information}\label{perfectcsi}
\vspace{-0.15cm}
We begin our experimental evaluation by quantifying the impact of various \gls{sinr} estimators on system performance, where the predicted current slot's normalized post-equalization \gls{sinr}, denoted as $\tilde{\gamma}_{\mathrm{post},t}^{(u)}$, serves as input to the \gls{rl} model, as described in Section~\ref{state-space}. Performance is evaluated in terms of the achieved mean \textit{\gls{se}} and mean \gls{bler}. In detail, when using the Oracle as input to \ariadne's \gls{rl} state, we rely on the ground-truth \gls{sinr} at the current slot, as reported by \sionnasys. When using one of the predictors, namely \gls{dt}~\cite{quinlan1986induction}, \gls{kf}~\cite{kalman1960}, \gls{rf}~\cite{breiman2001rf}, and \gls{oco}~\cite{maggi2026sinr}, we replace the Oracle with the predicted \gls{sinr} in order to assess how closely the resulting performance matches that of the Oracle, which serves as an upper bound. We then compare all the aforementioned \gls{rl}-based methods against 
state-of-the-art baselines, namely \gls{olla} and \gls{salad}. Finally, we also discard the current slot's predicted \gls{sinr} as input to the \gls{rl} model (denoted as \gls{dcqi} in the plots) and instead rely solely on past \gls{cqi} observations, evaluating the resulting performance across all methods.
\vspace{-0.4cm}
\begin{figure}[h]
    \centering
    \resizebox{\columnwidth}{!}{\input{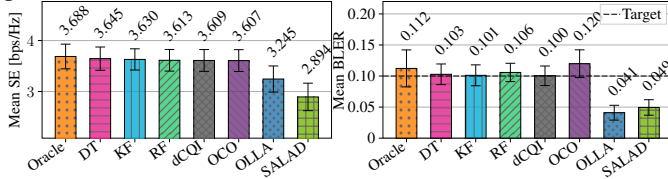}}
    \setlength{\abovecaptionskip}{-15pt}
    \setlength{\belowcaptionskip}{-2pt}
    \caption{Mean Spectral Efficiency and \gls{bler} for all \gls{rl} methods and baselines under Setup~A from Table~\ref{tab:configurations}.}
    \label{fig:predictors_summary}
    \vspace{-0.3cm}
\end{figure}

Regarding the inputs used by each \gls{sinr} predictor\footnote{We do not aim to directly compare the predictors. Instead, we assess whether the performance of the \gls{rl} agent is affected by the predictor choice, focusing on the robustness of the approach. The \gls{kf} is a constant-velocity filter with \gls{harq}-driven bias correction, adaptive process noise, and innovation gating. The \gls{dt} and \gls{rf} are trained online on past \gls{sinr} observations, and retrained every $N_\text{slots}$. The \gls{oco} follows the \ack/\nack-only feedback setting of~\cite{maggi2026sinr}, using mirror descent with Fixed-Share expert mixing over $12$ experts defined on the grid $\eta \in \{0.5, 1, 2, 3\}$, $\beta \in \{0, 0.15, 0.3\}$, with $\alpha = 0$.}, \gls{kf}, \gls{dt}, and \gls{rf} leverage past \gls{sinr} 
observations, \gls{kf} and \gls{oco} incorporate the most 
recent \gls{harq} feedback, while \gls{dt} and \gls{rf} rely on the 
\gls{harq} accumulator (i.e., $\Delta_{\mathrm{offset},t}^{(u)}$) to 
capture feedback trends. In contrast, in our implementation, 
\gls{oco} does not rely on explicit \gls{sinr} history and instead 
operates using only \gls{harq} outcomes and the previously selected 
\gls{mcs} index~\cite{maggi2026sinr}.

\vspace{-0.5cm}
\begin{table}[h]
\centering
\setlength{\abovecaptionskip}{-2pt}
\setlength{\belowcaptionskip}{-1pt}
\setlength{\tabcolsep}{4pt}
\caption{Performance comparison with respect to the Oracle.}
\label{tab:relative_performance}
\footnotesize
\begin{tabular}{@{}lccc@{}}
\toprule
\textbf{Method} & \textbf{SE [bps/Hz]} & \textbf{$\Delta$SE vs. Oracle [\%]} & \textbf{BLER} \\
\midrule
Oracle \gls{rl} & 3.688 & 0.0    & 0.112 \\
\gls{dt} \gls{rl}     & 3.645 & -1.17  & 0.103 \\
\gls{kf} \gls{rl}     & 3.630 & -1.57  & 0.101 \\
\gls{rf} \gls{rl}     & 3.613 & -2.03  & 0.106 \\
dCQI \gls{rl}   & 3.609 & -2.14  & 0.100 \\
\gls{oco} \gls{rl}    & 3.607 & -2.20  & 0.120 \\
\gls{olla}      & 3.245 & -12.01 & 0.041 \\
\gls{salad}     & 2.894 & -21.53 & 0.049 \\[-3pt]
\bottomrule
\end{tabular}
\vspace{-0.38cm}
\end{table}

In Fig.~\ref{fig:predictors_summary}, we observe that \ariadne, regardless of the 
predictor used, 
consistently outperforms \gls{olla} and 
\gls{salad}. Even without prediction of the current slot's \gls{sinr}, 
denoted as \gls{dcqi} in the plots, the \gls{rl} agent relying only on 
past \gls{cqi} reports outperforms \gls{olla} by $10\%$ in terms of 
spectral efficiency, achieving a mean spectral efficiency of 
$3.609$~bps/Hz compared to $3.245$~bps/Hz, at the cost of a higher mean 
\gls{bler} of $0.1$, whereas \gls{olla} achieves a \gls{bler} of $0.041$. Finally, \gls{olla} outperforms \gls{salad} by approximately $11\%$, 
while achieving similar \gls{bler} values. For the configuration of \gls{salad}, we use 
the default parameters as specified in~\cite{wiesmayr2025salad}.\footnote{\gls{salad} hyperparameters follow the reference implementation~\cite{nvlabs_salad}: learning rate $\varepsilon = 1.2$, bias score threshold $\rho = 0.25$, score window $T = 15$, probing probability $p^{\mathrm{probe}} = 0.15$, probing \gls{bler} target $\tau^{\mathrm{probe}} = 0.95$, and integral gain $k_E = 0.05$.} It is worth noting that \gls{salad} may require further parameter tuning for the specific scenario at hand, and therefore 
the parameterization used here may not be optimal for the current setup. 
This further highlights the advantage of \gls{rl} in adapting to changing 
channel conditions and dynamically adjusting its policy.
\vspace{-0.4cm}
\begin{figure}[h]
    \centering
    \resizebox{\columnwidth}{!}{\input{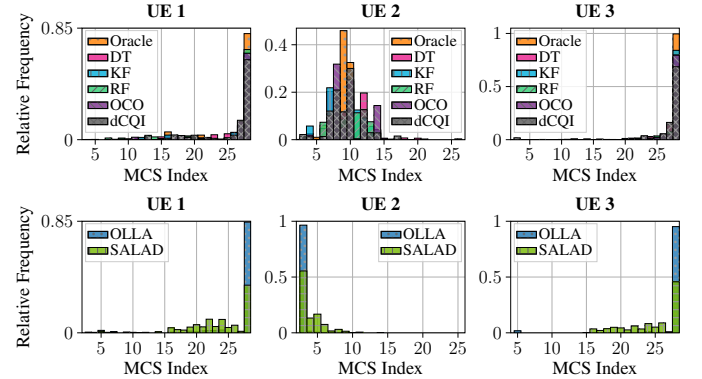}}
    \setlength{\abovecaptionskip}{-15pt}
    \setlength{\belowcaptionskip}{-5pt}
    \caption{Distribution of Selected \gls{mcs} Indices across Methods and \glspl{ue} under Setup~A from Table~\ref{tab:configurations}.}
    \label{fig:predictors_summary_2}
    \vspace{-0.4cm}
\end{figure}

In Table~\ref{tab:relative_performance}, we provide a summary of how closely all methods perform relative to the Oracle when used as input to \ariadne. We observe that \gls{dt} achieves a Spectral Efficiency closest to the Oracle, with only $\sim1\%$ lower performance, reporting a mean Spectral Efficiency of $3.645$~bps/Hz compared to $3.688$~bps/Hz for the 
Oracle, followed by \gls{kf}, whose Spectral Efficiency is reported at $3.630$~bps/Hz. All \gls{rl} methods achieve a Spectral Efficiency close to the Oracle, within $\sim2\%$. 
Finally, \gls{olla} reports a mean Spectral Efficiency that is 
$\sim12\%$ lower compared to the Oracle, while \gls{salad} underperforms all methods by $\sim20\%$. However, all \gls{rl} methods report a mean \gls{bler} of $0.1$, indicating the tradeoff between high throughput and 
reliability. In contrast, \gls{olla} and \gls{salad} maintain the \gls{bler} well below the $10\%$ target, at the cost of lower Spectral Efficiency.

Fig.~\ref{fig:predictors_summary_2} shows the relative frequency distribution of selected \gls{mcs} indices across different \gls{sinr} predictor methods and \glspl{ue}. Notably, the mean \gls{sinr} observed per \gls{ue} is $\sim26$~dB for \gls{ue}~$1$, $\sim6$~dB for \gls{ue}~$2$, and $\sim28$~dB for \gls{ue}~$3$. We observe that, for \gls{ue}~$1$ and \gls{ue}~$3$, all \gls{rl} methods select the highest \gls{mcs} indices, indicating agreement across methods in high-\gls{sinr} regimes. In contrast, for \gls{ue}~$2$, which experiences the lowest \gls{sinr}, the \gls{rl} agent adopts a more aggressive strategy, selecting \gls{mcs} indices between $5$ and $15$, with most selections 
concentrated around $8$--$12$. On the other hand, \gls{olla} and \gls{salad} predominantly select low \gls{mcs} indices, concentrated below $10$, indicating a more conservative approach.
\vspace{-0.25cm}
\subsection{Impact of the Adaptive BLER Penalty}\label{adapt-eval}
\begin{figure}[h]
    \centering
    \resizebox{\columnwidth}{!}{%
        \input{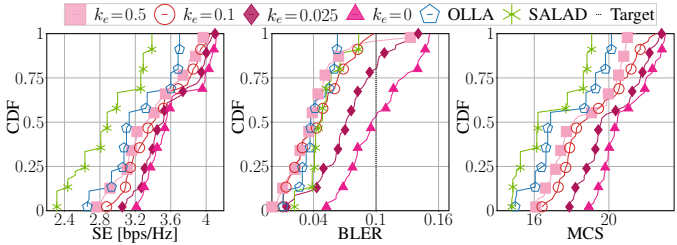}%
    }
    \setlength{\abovecaptionskip}{-15pt}
    \setlength{\belowcaptionskip}{-15pt}
    \caption{\glspl{cdf} of Spectral Efficiency, \gls{bler}, and \gls{mcs} under Setup~A from Table~\ref{tab:configurations}, leveraging the \gls{dt} predictor.}
    \label{fig:penalty_cdf_mcs}
    \vspace{-0.6cm}
\end{figure}

\vspace{-0.15cm}
We evaluate the impact of different $k_E$ values, as defined in~\eqref{eq:penaltyreward}--\eqref{eq:explicit-penalty-reward}. Since the previous results do not explicitly constrain the \gls{bler} and 
yield values above $10\%$ for all predictor modes and the Oracle, in contrast to \gls{salad} and \gls{olla}, which maintain \gls{bler} below $0.1$, we vary $k_E \in \{0,\,0.025,\,0.1,\,0.5\}$. Here, $k_E=0$ 
corresponds to no penalty, $k_E=0.025$ provides a moderate penalty, and $k_E=0.1$ and $k_E=0.5$ impose increasingly strong penalties on the median \gls{bler}. The resulting performance is summarized in Table~\ref{tab:ke_impact} and further illustrated in Fig.~\ref{fig:penalty_cdf_mcs}. As expected, 
increasing $k_E$ leads to more conservative \gls{mcs} selection, reducing Spectral Efficiency while improving reliability. In particular, $k_E=0$ achieves the highest Spectral Efficiency but at the cost of a higher 
\gls{bler}, while larger values of $k_E$ progressively reduce the median \gls{bler} at the expense of Spectral Efficiency. Among all configurations, $k_E=0.1$ provides the best tradeoff, achieving high Spectral Efficiency while consistently maintaining the \gls{bler} 
below the $10\%$ target (Fig.~\ref{fig:penalty_cdf_mcs}). Finally, \gls{olla} and \gls{salad} remain the most conservative.
\vspace{-0.5cm}

\begin{table}[h]
\centering
\setlength{\abovecaptionskip}{-4pt}
\setlength{\belowcaptionskip}{-9pt}
\caption{Impact of $k_E$ on Median Spectral Efficiency, \gls{bler}, and \gls{mcs}}
\label{tab:ke_impact}
\footnotesize
\begin{tabular}{@{}lccc@{}}
\toprule
$k_E$ & SE [bps/Hz] & BLER & MCS \\
\midrule
0     & 3.547 & 0.096 & 21 \\
0.025 & 3.520 & 0.072 & 20 \\
0.1   & 3.421 & 0.047 & 19 \\
0.5   & 3.342 & 0.037 & 18 \\
\midrule
OLLA  & 3.131 & 0.041 & 17 \\
SALAD & 2.882 & 0.049 & 16 \\
\bottomrule
\end{tabular}
\vspace{-0.4cm}
\end{table}

\begin{figure}[h]
    \centering
    \resizebox{\columnwidth}{!}{%
        \input{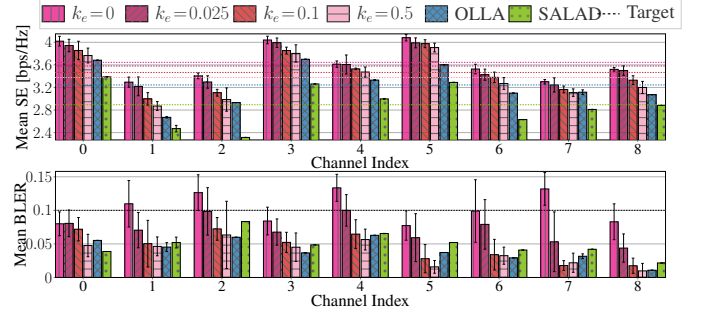}%
    }
    \setlength{\abovecaptionskip}{-15pt}
    \setlength{\belowcaptionskip}{-10pt}
    \caption{Spectral Efficiency and \gls{bler} under Setup~A from Table~\ref{tab:configurations}, using the \gls{dt} predictor across multiple channel realizations and experiments.}
    \label{fig:penalty_se_all}
    \vspace{-0.58cm}
\end{figure}

\vspace{-0.3cm}
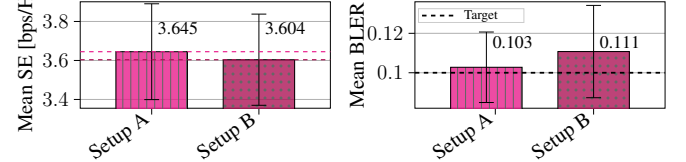
\begin{figure}[h]
    \centering
    \resizebox{0.98\columnwidth}{!}{%

\begin{tikzpicture}

\definecolor{darkgray176}{RGB}{176,176,176}
\definecolor{lightgray204}{RGB}{204,204,204}
\definecolor{steelblue31119180}{RGB}{31,119,180}
\definecolor{hotpinkE91E8C}{RGB}{233,30,140}
\definecolor{keBerryAD1457}{RGB}{173,20,87}

\begin{groupplot}[
  group style={group size=2 by 1, horizontal sep=2cm},
  width=7.5cm,
  height=4.1cm,
  tick label style={font=\Large},
  label style={font=\Large},
  title style={font=\Large},
  xticklabel style={rotate=35, anchor=east, font=\Large},
  ylabel style={yshift=-0.1cm},
  scaled y ticks=false,
]

\nextgroupplot[
  tick align=outside, tick pos=left,
  x grid style={darkgray176}, y grid style={darkgray176},
  xmin=-0.4, xmax=1.0,
  xtick={0, 0.6},
  xticklabels={Setup A, Setup B},
  xtick style={color=black}, ytick style={color=black},
  ymajorgrids,
  ymin=3.355, ymax=3.901,
  ylabel={Mean SE [bps/Hz]},
]
\draw[draw=none, fill=hotpinkE91E8C, fill opacity=0.8] (axis cs:-0.2,2.000) rectangle (axis cs:0.2,3.64498);
\draw[draw=black, fill=none, pattern=vertical lines, pattern color=black!60] (axis cs:-0.2,2.000) rectangle (axis cs:0.2,3.64498);
\draw[draw=none, fill=keBerryAD1457, fill opacity=0.8] (axis cs:0.4,2.000) rectangle (axis cs:0.8,3.60388);
\draw[draw=black, fill=none, pattern=dots, pattern color=black!60] (axis cs:0.4,2.000) rectangle (axis cs:0.8,3.60388);
\path [draw=black, line width=0.56pt] (axis cs:0,3.39924) -- (axis cs:0,3.89072);
\path [draw=black, line width=0.56pt] (axis cs:0.6,3.37002) -- (axis cs:0.6,3.83775);
\addplot [forget plot, semithick, steelblue31119180, mark=-, mark size=5, mark options={solid,draw=black}, only marks] table {%
0   3.39924
0.6 3.37002
};
\addplot [forget plot, semithick, steelblue31119180, mark=-, mark size=5, mark options={solid,draw=black}, only marks] table {%
0   3.89072
0.6 3.83775
};
\addplot[hotpinkE91E8C, dashed, line width=0.8pt, forget plot] coordinates {(-0.4,3.64498) (1.0,3.64498)};
\addplot[keBerryAD1457, dashed, line width=0.8pt, forget plot] coordinates {(-0.4,3.60388) (1.0,3.60388)};
\draw (axis cs:0,3.7) node[scale=1.2, anchor=south west, text=black]{3.645};
\draw (axis cs:0.6,3.7) node[scale=1.2, anchor=south west, text=black]{3.604};

\nextgroupplot[
  tick align=outside, tick pos=left,
  x grid style={darkgray176}, y grid style={darkgray176},
  xmin=-0.4, xmax=1.0,
  xtick={0.05, 0.6},
  xticklabels={Setup A, Setup B},
  xtick style={color=black}, ytick style={color=black},
  ymajorgrids,
  ymin=0.082, ymax=0.136,
  ylabel={Mean BLER},
  ylabel style={yshift=-2pt},
]
\draw[draw=none, fill=hotpinkE91E8C, fill opacity=0.8] (axis cs:-0.2,0.000) rectangle (axis cs:0.2,0.102778);
\draw[draw=black, fill=none, pattern=vertical lines, pattern color=black!60] (axis cs:-0.2,0.000) rectangle (axis cs:0.2,0.102778);
\draw[draw=none, fill=keBerryAD1457, fill opacity=0.8] (axis cs:0.4,0.000) rectangle (axis cs:0.8,0.110746);
\draw[draw=black, fill=none, pattern=dots, pattern color=black!60] (axis cs:0.4,0.000) rectangle (axis cs:0.8,0.110746);
\path [draw=black, line width=0.56pt] (axis cs:0,0.0848971) -- (axis cs:0,0.120659);
\path [draw=black, line width=0.56pt] (axis cs:0.6,0.0873295) -- (axis cs:0.6,0.134162);
\addplot [forget plot, semithick, steelblue31119180, mark=-, mark size=5, mark options={solid,draw=black}, only marks] table {%
0   0.0848971
0.6 0.0873295
};
\addplot [forget plot, semithick, steelblue31119180, mark=-, mark size=5, mark options={solid,draw=black}, only marks] table {%
0   0.120659
0.6 0.134162
};
\addplot [forget plot, black, dashed, line width=1.2pt] coordinates {(-0.4,0.1) (1.0,0.1)};
\draw (axis cs:0,0.109) node[scale=1.2, anchor=south west, text=black]{0.103};
\draw (axis cs:0.6,0.109) node[scale=1.2, anchor=south west, text=black]{0.111};
\draw[draw=lightgray204, fill=white, fill opacity=0.8]
  (axis cs:-0.38,0.1255) rectangle (axis cs:0.21,0.1325);
\draw[black, dashed, line width=1.2pt]
  (axis cs:-0.36,0.129) -- (axis cs:-0.18,0.129);
\draw (axis cs:-0.15,0.129) node[scale=0.9, anchor=west, text=black]{\normalsize Target};

\end{groupplot}

\end{tikzpicture}%
    }
    \setlength{\abovecaptionskip}{-6.5pt}
    \setlength{\belowcaptionskip}{-12pt}
    \caption{Mean Spectral Efficiency and \gls{bler} for all \gls{rl} methods and baselines under Setup~A and~B from Table~\ref{tab:configurations}, leveraging the \gls{dt} predictor.}
    \label{fig:cqiSize_comparison}
    \vspace{-0.4cm}
\end{figure}

Finally, in Fig.~\ref{fig:penalty_se_all}, we observe the mean Spectral Efficiency across each channel realization, averaged over multiple experiments. We observe that \gls{rl} with $k_E=0.1$ consistently meets the $10\%$ \gls{bler} target while achieving one of the highest Spectral 
Efficiency values among all considered methods, outperforming the industry-standard \gls{olla}. In contrast, \gls{rl} with $k_E=0$ and $k_E=0.025$ achieves higher Spectral Efficiency but often reaches or 
exceeds the \gls{bler} target. Based on these results, we identify $k_E = 0.1$ as the recommended operating point, as it consistently meets the $10\%$ \gls{bler} target while achieving high Spectral Efficiency outperforming both \gls{olla} and \gls{salad}.
\vspace{-0.25cm}
\subsection{Impact of Observation Window Size}\label{obssize-eval}

\vspace{-0.15cm}
We now proceed by evaluating how different observation window sizes impact the mean performance of the \gls{rl} agent. Once again, for the current slot's \gls{sinr} prediction, we leverage the \gls{dt}, as it 
achieved the closest performance to the Oracle in terms of Spectral Efficiency in the previous evaluation. The parameterization for this step 
is given in Table~\ref{tab:configurations}. As shown in Fig.~\ref{fig:cqiSize_comparison}, a smaller observation window 
(Setup~B) results in a slightly lower mean Spectral Efficiency of $3.604$~bps/Hz compared to $3.645$~bps/Hz achieved by Setup~A. Although both Spectral Efficiency values are comparable, Setup~A achieves a lower \gls{bler} 
($0.103$) than Setup~B ($0.111$). Both values remain above the $10\%$ \gls{bler} target, which is expected since the adaptive penalty reward is 
not leveraged in these experiments. Overall, Setup~A is preferred, as it achieves slightly higher Spectral 
Efficiency while maintaining a lower \gls{bler}, enabling the \gls{rl} agent to select actions without significantly exceeding the \gls{bler} target.
\vspace{-0.2cm}
\section{Site-Specific Training and Robustness Across Channel Scenarios}
\vspace{-0.15cm}
\begin{figure}[h]
    \centering
    \resizebox{0.99\columnwidth}{!}{%
        \input{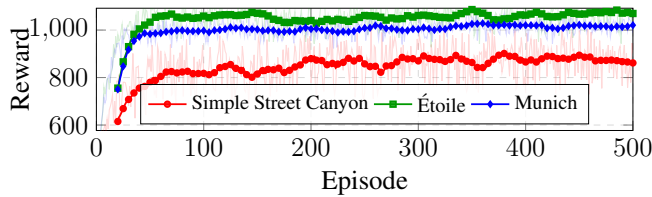}%
    }
    \setlength{\abovecaptionskip}{-18pt}
    \setlength{\belowcaptionskip}{-15pt}
    \caption{Training reward over episodes.}
    \label{fig:training_results}
    \vspace{-0.65cm}
\end{figure}
Fig.~\ref{fig:training_results} shows the training reward evolution across three different channel scenarios in \sionna. The \gls{rl} 
agent consistently improves its performance and converges in all considered environments. This indicates that the learned policy adapts to varying channel characteristics.
\vspace{-0.25cm}
\section{Future Integration of \gls{rl} on \gls{ota} \gls{5g} Testbeds}\label{nextsteps}
\begin{figure}[h]
    \centering
    \resizebox{0.95\columnwidth}{!}{%

\begin{tikzpicture}

\definecolor{bellmanblue}{RGB}{31,119,180}
\definecolor{schedblue}{RGB}{100,149,200}
\definecolor{fqicoral}{RGB}{250,160,120}

\begin{groupplot}[
  group style={
    group size=2 by 1,
    horizontal sep=1.6cm,
  },
  height=3.5cm,
  tick align=outside,
  tick pos=left,
  xmajorgrids,
  ymajorgrids,
  x grid style={black!15},
  y grid style={black!15},
  tick label style={font=\Large},
  label style={font=\Large},
]

\nextgroupplot[
  width=5cm,
  ylabel style={yshift=-0.2cm},
  xlabel={FQI Iteration},
  ylabel={Avg.\ Q-value},
  xmin=0, xmax=30,
  ymin=60, ymax=180,
  legend style={at={(0.97,0.35)}, anchor=east, font=\Large},
]

\addplot [line width=1.5pt, bellmanblue, mark=*, mark size=1.8pt]
table {%
0  79.41
1  120.15
2  140.77
3  150.99
4  156.21
5  158.79
6  160.13
7  160.74
8  161.09
9  161.21
10 161.28
11 161.33
12 161.31
13 161.35
14 161.31
15 161.35
16 161.34
17 161.34
18 161.36
19 161.38
20 161.37
21 161.34
22 161.40
23 161.37
24 161.38
25 161.38
26 161.41
27 161.33
28 161.31
29 161.30
30 161.25
};

\nextgroupplot[
  ybar,
  bar width=8.5pt,
  width=12cm,
  ylabel style={yshift=-0.2cm},
  xlabel={MCS Index},
  ylabel={Percentage [\%]},
  xtick={11,12,13,14,15,16,17,18,19,20,21,22},
  xticklabels={11,12,13,14,15,16,17,18,19,20,21,22},
  x tick label style={rotate=45, anchor=east, font=\Large},
  xmin=10, xmax=22.8,
  ymin=0, ymax=75,
  legend style={at={(0.03,0.97)}, anchor=north west, font=\Large},
  xmajorgrids=false,
]

\addplot [fill=schedblue, fill opacity=0.75, draw=schedblue!80!black]
coordinates {
  (0,  0.0)
  (1,  0.0)
  (2,  0.0)
  (3,  0.0)
  (4,  0.0)
  (5,  0.0)
  (6,  0.0)
  (7,  0.0)
  (8,  0.0)
  (9,  0.0)
  (10,  0.0)
  (11,  3.6)
  (12,  0.0)
  (13,  1.8)
  (14,  3.6)
  (15,  1.8)
  (16,  1.8)
  (17,  5.4)
  (18,  8.9)
  (19,  71.4)
  (20,  0.0)
  (21,  1.8)
  (22,  0.0)
};
\addlegendentry{OAI Scheduler}

\addplot [
  fill=fqicoral,
  fill opacity=1,
  draw=fqicoral!80!black,
  postaction={pattern=north east lines, pattern color=fqicoral!80!black}
]
coordinates {
  (0,  0.0)
  (1,  0.0)
  (2,  0.0)
  (3,  0.0)
  (4,  0.0)
  (5,  0.0)
  (6,  0.0)
  (7,  0.0)
  (8,  0.0)
  (9,  0.0)
  (10,  0.0)
  (11,  0.0)
  (12,  0.0)
  (13,  0.0)
  (14,  0.0)
  (15,  3.6)
  (16,  1.8)
  (17,  25.0)
  (18,  3.6)
  (19,  35.7)
  (20,  1.8)
  (21,  26.8)
  (22,  1.8)
};
\addlegendentry{FQI Policy}

\end{groupplot}
\end{tikzpicture}%
    }

    \vspace{-0.25cm}

    \resizebox{0.95\columnwidth}{!}{%

\begin{tikzpicture}

\definecolor{bellmanblue}{RGB}{31,119,180}
\definecolor{schedblue}{RGB}{100,149,200}
\definecolor{fqicoral}{RGB}{250,160,120}

\begin{groupplot}[
  group style={
    group size=2 by 1,
    horizontal sep=1.8cm,
  },
  height=3.5cm,
  tick align=outside,
  tick pos=left,
  ymajorgrids,
  y grid style={black!15},
  tick label style={font=\Large},
  label style={font=\Large},
]

\nextgroupplot[
  width=5cm,
  ylabel style={yshift=-0.2cm},
  xlabel={FQI Iteration},
  ylabel={Avg.\ Q-value},
  xmajorgrids,
  x grid style={black!15},
  xmin=0, xmax=30,
  ymin=700, ymax=1670,
  ytick={800,1200,1600},
  legend style={at={(0.97,0.25)}, anchor=east, font=\Large},
]

\addplot [line width=1.5pt, bellmanblue, mark=*, mark size=1.8pt]
table {%
0  754.10
1  1200.96
2  1425.94
3  1538.28
4  1595.28
5  1623.44
6  1637.97
7  1644.87
8  1648.54
9  1649.37
10  1649.71
11  1651.14
12  1650.57
13  1650.88
14  1650.74
15  1651.49
16  1652.05
17  1651.20
18  1650.65
19  1650.89
20  1651.31
21  1650.78
22  1650.95
23  1651.25
24  1650.52
25  1650.35
26  1650.87
27  1650.55
28  1650.35
29  1651.01
30  1651.13
};

\nextgroupplot[
  ybar,
  bar width=8.5pt,
  width=12cm,
  ylabel style={yshift=-0.2cm},
  xlabel={MCS Index},
  ylabel={Percentage [\%]},
  xtick={9,10,11,12,13,14,15,16,17,18,19,20,21,22,23,24,25,26,27},
  xticklabels={9,10,11,12,13,14,15,16,17,18,19,20,21,22,23,24,25,26,27},
  x tick label style={rotate=45, anchor=east, font=\Large},
  xmin=8, xmax=27.8,
  ymin=0, ymax=76,
  legend style={at={(0.05,0.97)}, anchor=north west, font=\Large},
  xmajorgrids=false,
]

\addplot [fill=schedblue, fill opacity=0.75, draw=schedblue!80!black]
coordinates {
  (9,  0.17)
  (10,  0.11)
  (11,  0.10)
  (12,  0.09)
  (13,  0.20)
  (14,  0.12)
  (15,  0.24)
  (16,  0.10)
  (17,  0.19)
  (18,  0.14)
  (19,  0.19)
  (20,  0.37)
  (21,  1.21)
  (22,  1.05)
  (23,  16.91)
  (24,  2.49)
  (25,  10.79)
  (26,  3.00)
  (27,  62.54)
};
\addlegendentry{OAI Scheduler}

\addplot [
  fill=fqicoral,
  fill opacity=1,
  draw=fqicoral!80!black,
  postaction={pattern=north east lines, pattern color=fqicoral!80!black}
]
coordinates {
  (9,  0.70)
  (10,  0.00)
  (11,  5.40)
  (12,  0.10)
  (13,  0.00)
  (14,  17.40)
  (15,  2.00)
  (16,  0.00)
  (17,  0.00)
  (18,  7.80)
  (19,  0.00)
  (20,  16.10)
  (21,  4.10)
  (22,  0.00)
  (23,  19.20)
  (24,  0.70)
  (25,  0.50)
  (26,  0.00)
  (27,  26.00)
};
\addlegendentry{FQI Policy}

\end{groupplot}
\end{tikzpicture}%
    }
    \setlength{\abovecaptionskip}{-2.5pt}
    \setlength{\belowcaptionskip}{-5pt}
    \caption{Offline \gls{fqi} results on \gls{ota} data for \gls{ul} (top) and \gls{dl} (bottom). Left: Q-value convergence across Bellman iterations. Right: \gls{mcs} distribution of the learned policy vs.\ the \gls{oai} scheduler.}
    \label{fig:fqi}
    \vspace{-0.3cm}
\end{figure}
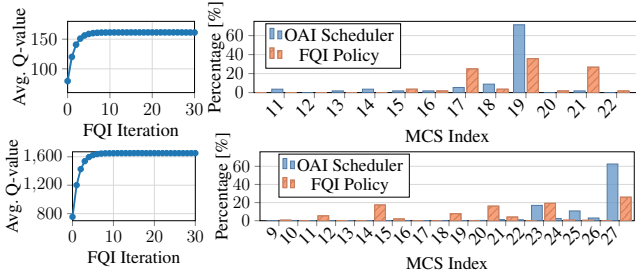

 \vspace{-0.15cm}
To examine whether reward-driven \gls{mcs} selection extends beyond simulation, we train an \gls{fqi} agent~\cite{ernst2005tree} on \gls{ota} measurements collected from a \gls{5g} testbed~\cite{villa2025x5g}, using both \gls{dl} and \gls{ul} data. The state consists of \gls{cqi}, \gls{rsrp}, and instantaneous \gls{bler}; the reward is a proxy for achieved throughput derived from logged measurements; and the Q-function is an \gls{rf} with $\gamma = 0.5$. As shown in Fig.~\ref{fig:fqi}, Q-function targets stabilize across Bellman iterations, and the learned policy shifts \gls{mcs} selections away from the scheduler's preferred indices. Since these datasets reflect only the \gls{oai} \gls{olla} scheduler's choices, closed-loop evaluation is left for future work. The fact that both this offline agent and the online \gls{ppo} agent independently deviate from \gls{olla} suggests that reward-driven \gls{mcs} selection tends to identify operating points that rule-based schedulers do not.
\vspace{-0.5cm}
 \vspace{-0.35cm}
\section{Conclusions and Future Work}\label{conclusion} 
\vspace{-0.15cm}
We developed \ariadne, a framework that seamlessly integrates with \sionna, leveraging online \gls{rl} for \gls{mcs} selection. The use of system-level simulators, which enable the on-the-fly integration 
of \gls{ai} modules, has facilitated the exploration of various approaches, spanning reward formulation and observation window design, to study the impact of different design choices on link adaptation. Although the \sionna integration provides a practical first step toward 
the design of \gls{rl}-based link adaptation, future work will focus on the deployment of \ariadne on \gls{ota} \gls{5g} testbeds. Preliminary results on both \gls{dl} and \gls{ul} \gls{5g} data demonstrate 
the potential of learning-based approaches in real-world scenarios.
\vspace{-0.2cm}

\vspace{-0.2cm}
\bibliographystyle{IEEEtran}
\bibliography{IEEEabrv,ref}

\begin{thebibliography}{10}
\providecommand{\url}[1]{#1}
\csname url@samestyle\endcsname
\providecommand{\newblock}{\relax}
\providecommand{\bibinfo}[2]{#2}
\providecommand{\BIBentrySTDinterwordspacing}{\spaceskip=0pt\relax}
\providecommand{\BIBentryALTinterwordstretchfactor}{4}
\providecommand{\BIBentryALTinterwordspacing}{\spaceskip=\fontdimen2\font plus
\BIBentryALTinterwordstretchfactor\fontdimen3\font minus \fontdimen4\font\relax}
\providecommand{\BIBforeignlanguage}[2]{{%
\expandafter\ifx\csname l@#1\endcsname\relax
\typeout{** WARNING: IEEEtran.bst: No hyphenation pattern has been}%
\typeout{** loaded for the language `#1'. Using the pattern for}%
\typeout{** the default language instead.}%
\else
\language=\csname l@#1\endcsname
\fi
#2}}
\providecommand{\BIBdecl}{\relax}
\BIBdecl

\bibitem{nvidia_nokia2025}
{NVIDIA Corporation}, ``{NVIDIA and Nokia to Pioneer the {AI} Platform for {6G} --- Powering America's Return to Telecommunications Leadership},'' NVIDIA Newsroom, Press Release, Oct. 2025, [Available Online]: \url{https://nvidianews.nvidia.com/ news/nvidia-nokia-ai-telecommunications}.

\bibitem{nokia_ran_dt2026}
{Nokia}, ``{Nokia launches {Nokia RAN Digital Twin} to turbo-charge {AI}-native {6G}, powered by {NVIDIA Aerial Omniverse Digital Twin}},'' Nokia Corporate Blog, Feb. 2026, [Available Online]: \url{https://www.nokia.com/blog/ nokia-launches-nokia-ran-digital-twin-to-turbo-charge- ai-native-6g-powered-by-nvidia-aerial-omniverse- digital-twin/}.

\bibitem{polese2024colosseum}
M.~Polese, L.~Bonati, S.~D’Oro, P.~Johari, D.~Villa, S.~Velumani, R.~Gangula, M.~Tsampazi, C.~P. Robinson, G.~Gemmi \emph{et~al.}, ``{Colosseum: The open RAN digital twin},'' \emph{IEEE Open Journal of the Communications Society}, 2024.

\bibitem{sionna2022}
J.~Hoydis, S.~Cammerer, F.~Ait~Aoudia, M.~Nimier-David, A.~Keller, A.~Vem, M.~Stark, and T.~O’Shea, ``{Sionna: An Open-Source Library for Next-Generation Physical Layer Research},'' \emph{arXiv preprint arXiv:2203.11854}, 2022.

\bibitem{khedhri2025adaptive}
N.~Khedhri and M.~Najar, ``Adaptive modulation selection in wireless communications: A comparative study of reinforcement learning, deep learning, deep reinforcement learning, and traditional policies,'' in \emph{International Wireless Communications and Mobile Computing (IWCMC)}.\hskip 1em plus 0.5em minus 0.4em\relax IEEE, 2025, pp. 1622--1625.

\bibitem{peri2025offline}
S.~Peri, A.~Russo, G.~Fodor, and P.~Soldati, ``{Offline reinforcement learning and sequence modeling for downlink link adaptation},'' in \emph{International Conference on Machine Learning for Communication and Networking (ICMLCN)}.\hskip 1em plus 0.5em minus 0.4em\relax IEEE, 2025, pp. 1--7.

\bibitem{pulliyakode2017reinforcement}
S.~K. Pulliyakode and S.~Kalyani, ``{Reinforcement learning techniques for outer loop link adaptation in 4G/5G systems},'' \emph{arXiv preprint arXiv:1708.00994}, 2017.

\bibitem{saxena2019contextual}
V.~Saxena, J.~Jald{\'e}n, J.~E. Gonzalez, M.~Bengtsson, H.~Tullberg, and I.~Stoica, ``{Contextual multi-armed bandits for link adaptation in cellular networks},'' in \emph{Workshop on Network Meets AI \& ML}, 2019, pp. 44--49.

\bibitem{zubow2021grgym}
A.~Zubow, S.~R{\"o}sler, P.~Gaw{\l}owicz, and F.~Dressler, ``{GrGym: When GNU radio goes to (AI) gym},'' in \emph{Proc. of the 22nd International Workshop on Mobile Computing Systems and Applications}, 2021, pp. 8--14.

\bibitem{leite2012flexible}
J.~P. Leite, P.~H.~P. de~Carvalho, and R.~D. Vieira, ``{A flexible framework based on reinforcement learning for adaptive modulation and coding in OFDM wireless systems},'' in \emph{Wireless Communications and Networking Conference (WCNC)}.\hskip 1em plus 0.5em minus 0.4em\relax IEEE, 2012, pp. 809--814.

\bibitem{pedersen2007frequency}
K.~I. Pedersen, G.~Monghal, I.~Z. Kovacs, T.~E. Kolding, A.~Pokhariyal, F.~Frederiksen, and P.~Mogensen, ``{Frequency domain scheduling for OFDMA with limited and noisy channel feedback},'' in \emph{IEEE 66th Vehicular Technology Conference}, 2007, pp. 1792--1796.

\bibitem{kela2022reinforcement}
P.~Kela, T.~H{\"o}hne, T.~Veijalainen, and H.~Abdulrahman, ``Reinforcement learning for delay sensitive uplink outer-loop link adaptation,'' in \emph{Joint European Conference on Networks and Communications \& 6G Summit}.\hskip 1em plus 0.5em minus 0.4em\relax IEEE, 2022, pp. 59--64.

\bibitem{wiesmayr2025salad}
R.~Wiesmayr, L.~Maggi, S.~Cammerer, J.~Hoydis, F.~A. Aoudia, and A.~Keller, ``{SALAD: Self-adaptive link adaptation},'' \emph{arXiv preprint arXiv:2510.05784}, 2025.

\bibitem{maggi2026sinr}
L.~Maggi, B.~Bonev, R.~Wiesmayr, S.~Cammerer, and A.~Keller, ``{SINR Estimation under Limited Feedback via Online Convex Optimization},'' \emph{arXiv preprint arXiv:2603.02061}, 2026.

\bibitem{3gpp38214}
{3rd Generation Partnership Project}, ``Nr; physical layer procedures for data (3gpp ts 38.214),'' 2023.

\bibitem{schulman2017proximal}
J.~Schulman, F.~Wolski, P.~Dhariwal, A.~Radford, and O.~Klimov, ``{Proximal Policy Optimization Algorithms},'' 2017, [Available Online]: \url{https://arxiv.org/abs/1707.06347}.

\bibitem{stable-baselines3}
A.~Raffin, A.~Hill, A.~Gleave, A.~Kanervisto, M.~Ernestus, and N.~Dormann, ``{Stable-Baselines3: Reliable Reinforcement Learning Implementations},'' GitHub Repository, 2021, [Available Online]: \url{https://github.com/DLR-RM/stable-baselines3}.

\bibitem{gymnasium2023}
{Farama Foundation}, ``{Gymnasium: A Standard Interface for Reinforcement Learning Environments},'' GitHub Repository, 2023, [Available Online]: \url{https://github.com/Farama-Foundation/Gymnasium}.

\bibitem{quinlan1986induction}
J.~R. Quinlan, ``{Induction of decision trees},'' \emph{Machine learning}, vol.~1, no.~1, pp. 81--106, 1986.

\bibitem{kalman1960}
R.~E. Kalman, ``{A New Approach to Linear Filtering and Prediction Problems},'' \emph{Transactions of the ASME--Journal of Basic Engineering}, vol.~82, pp. 35--45, 1960.

\bibitem{breiman2001rf}
L.~Breiman, ``{Random Forests},'' \emph{Machine Learning}, vol.~45, 2001.

\bibitem{nvlabs_salad}
{NVlabs}, ``{SALAD: Self-Adaptive Link Adaptation --- Reference Implementation},'' GitHub Repository, NVlabs, 2025, [Available Online]: \url{https://github.com/NVlabs/salad/blob/ main/notebooks/meet_salad.ipynb}.

\bibitem{ernst2005tree}
D.~Ernst, P.~Geurts, and L.~Wehenkel, ``{Tree-based batch mode reinforcement learning},'' \emph{Journal of Machine Learning Research}, vol.~6, 2005.

\bibitem{villa2025x5g}
D.~Villa, I.~Khan, F.~Kaltenberger, N.~Hedberg, R.~S. da~Silva, S.~Maxenti, L.~Bonati, A.~Kelkar, C.~Dick, E.~Baena \emph{et~al.}, ``{X5G: An open, programmable, multi-vendor, end-to-end, private 5G O-RAN testbed with NVIDIA ARC and OpenAirInterface},'' \emph{IEEE Transactions on Mobile Computing}, 2025.

\end{thebibliography}

\end{document}